\documentclass[a4paper,12pt]{article}
\usepackage{amsmath}
\usepackage{amsfonts}
\usepackage{amssymb}
\usepackage[utf8]{inputenc}
\pdfoutput=1
\usepackage{graphicx}
\usepackage{color}
\usepackage{braket}
\usepackage{dcolumn}
\usepackage{bm,url}
\usepackage[linktocpage]{hyperref}
\usepackage{subfigure}
\usepackage{amsfonts}
\usepackage{setspace}
\usepackage[usenames,dvipsnames,svgnames]{xcolor}  
\usepackage{hyperref}   
\definecolor{caribbeangreen}{rgb}{0.0, 0.8, 0.6}
\definecolor{darkpastelgreen}{rgb}{0.01, 0.75, 0.24}
\definecolor{darkturquoise}{rgb}{0.0, 0.61, 0.72}
\hypersetup{
    breaklinks=true,
    pdfstartview={FitH},    
    colorlinks=true,       
    linkcolor=Magenta,          
    citecolor=Blue,        
    filecolor=Magenta,      
    urlcolor=darkturquoise,           
    anchorcolor=green,      
    linktocpage=true
}
\usepackage[left=2cm,right=2cm,top=3cm,bottom=3.5cm]{geometry}

\def\vhrulefill#1{\leavevmode\leaders\hrule\@height#1\hfill \kern\z@}
\makeatother
\begin{document}
\newcommand\be{\begin{equation}}
\newcommand\ee{\end{equation}}
\newcommand\bea{\begin{eqnarray}}
\newcommand\eea{\end{eqnarray}}
\newcommand\bseq{\begin{subequations}} 
\newcommand\eseq{\end{subequations}}
\newcommand\bcas{\begin{cases}}
\newcommand\ecas{\end{cases}}
\newcommand{\p}{\partial}
\newcommand{\f}{\frac}
\newcommand{\red}{\textcolor{red}}
\newcommand{\blue}{\textcolor{blue}}
\newcommand{\emailadd}[1]{{\centering\parbox[t]{1\textwidth}
{\centering#1}}\vspace{0.4cm}}
\vspace{5mm}
\vspace{0.5cm}

\begin{center}
\def\thefootnote{\fnsymbol{footnote}}
{\bf \large  Harvesting energy driven by Comisso--Asenjo process from Kerr--MOG black holes
}\\  [0.2cm]

\vspace{1cm}
{ \bf \large Mohsen Khodadi,$^{a,b}$}
{\bf \large David F. Mota,$^{c}$}
{ \bf \large Ahmad Sheykhi$^{a,b}$}
\\[0.7cm]

{$^a$ Department of Physics, College of Sciences, Shiraz University, Shiraz 71454, Iran}\\

{$^b$ Biruni Observatory, College of Sciences, Shiraz University, Shiraz 71454, Iran}\\

{$^c$ Institute of Theoretical Astrophysics, University of Oslo, P.O. Box 1029 Blindern, N-0315 Oslo,
Norway}\\

\vspace{1cm}
\emailadd {E-mail:
\blue {m.khodadi@hafez.shirazu.ac.ir},
\blue{d.f.mota@astro.uio.no},
\blue{asheykhi@shirazu.ac.ir}}
\end{center}
\vspace{.001cm}


\vspace{.1cm}
\makeatletter
\def\vhrulefill#1{\leavevmode\leaders\hrule\@height#1\hfill \kern\z@}
\makeatother
\setcounter{footnote}{0}

\begin{abstract}
Magnetic reconnection is a process that plays a critical role in
plasma astrophysics by converting magnetic energy into plasma
particle energy. Recently, Comisso and Asenjo demonstrated that
rapid magnetic reconnection within a black hole's ergosphere can
efficiently extract energy from a rotating black hole.  In this
paper, by considering a Kerr black hole in the MOdified gravity
(MOG) framework, we investigate the impact of the MOG parameter
$\alpha$ on the rotational energy extraction via the
Comisso--Asenjo process (CAP). To model energy extraction from supermassive black holes located in the center of galaxies,  we set the value of $\alpha$ within the range inferred from the recent observation of Sgr A* by the Event Horizon Telescope (EHT).  Our
results indicate that the Kerr--MOG black hole is a more efficient
host for CAP--based rotational energy extraction compared to the
Kerr black hole, since it amplifies the power of energy extraction
and efficiency of the plasma energization process. We show that,
from the energy extraction viewpoint, the CAP is more efficient
than the Blandford--Znajek process (BZP). The latter is another
magnetic field--based energy extraction model which is widely
believed to be an engine for powering the high--energy
astrophysics jets emerging from the supermassive black holes at
active galactic nuclei. In particular, we show that the ratio of
the energy extraction power of CAP to BZP in the presence of the
MOG parameter is greater than that of the Kerr black hole. Our
results promise this phenomenological message that the
MOG--induced correction on the Kerr black hole background plays an
important role in favor of energy extraction via the CAP.
\\\\

\end{abstract}
\newpage
\vspace{0.5cm} \vhrulefill{1.2pt}
\tableofcontents
\vspace{0.5cm} \vhrulefill{1.2pt}

\section{Introduction} \label{Intro}
It is a general belief that black holes are among the most
controversial and remarkable predictions of Einstein's theory of
general relativity. In recent years, the advent of advanced
instruments such as gravitational wave interferometer
\cite{LIGOScientific:2007fwp,Accadia,KAGRA:2013rdx} and EHT
\cite{EventHorizonTelescope:2019uob} have added to the
attractiveness of this topic, as they have opened novel avenues
for the phenomenological survey of black hole physics. Finding the source of relativistic jets and beams received from the center of active galactic nuclei (AGNs) is one of the highly regarded and in--progress subjects to research in the high--energy astrophysics \cite{book}. Black holes are expected to be viable candidates to play the role of the powerful engines of energy, meaning that one of the most promising processes prone to producing relativistic jets is rotational energy extraction from a spinning black hole
\footnote{Regarding the extraction of rotational energy from a
black hole, it is useful to note that less than one--third of the
rest energy of a rotating black hole has the capability of
extraction, according to the seminal paper of Christodoulou
\cite{Christodoulou:1970wf}.} \cite{Misner,Bardeen:1972fi}. Black holes as natural particle accelerators provide a powerful tools to examine many branches of physics \cite{Banados:2009pr}. Namely, one can consider this compact object as generator
of substantial energy outflows released by some well--known
high--energy astrophysical systems, such as gamma--ray bursts
(GRBs) \footnote{Energy injection through neutrino annihilation
around black holes can be effective in explaining GRBs
\cite{Kumar:2014upa,Lambiase:2020pkc,Kimura:2022zyg,Khodadi:2023yiw}.
Note that this is not the case for AGN. }  \cite{Kulkarni:1999aa},
microquasar \cite{Mirabel:1994rb}, and relativistic jets powered
by supermassive black holes hosted in AGNs. 
These astronomical
systems, in essence, release collimated flows of plasma with
velocities around the speed of light \cite{Davis:2020wea}. Roughly speaking, it is believed that the gravitational potential energy of the matter falling toward a black hole during accretion, and electromagnetic field energy created in the background of a black hole (or even the black hole itself), can be high--energy jet--producing origins. In this regard, so far, various scenarios
have been put forward for theorizing about the processes related
to energy extraction from black holes. The main purpose of these scenarios is to physically justify this amazing phenomenon to us by revealing the physics behind black hole energy extraction.

Although it is not yet clear which process is responsible for the
production of high--energy jets, several well--known scenarios
have been proposed. The most well-known are the Penrose process
(PP) \cite{Penrose:1969pc,Penrose:1971uk},  the magnetic Penrose
process (MPP) \cite{Wagh:1985vuj}, superradiant scattering
\cite{Teukolsky:1974yv}, and the Blandford–Znajek process (BZP)
\cite{Blandford:1977ds} (see \cite{Komissarov:2004ms} for more).
These processes provide different backgrounds to address this phenomenon. Although the possibility of extracting energy from spinning black holes dates back historically to PP, from an astrophysics viewpoint, it is not desirable as it does not result in the significant extraction of rotational energy \footnote{In opposite to original PP its magnetic
version i.e., MPP is considerably efficient so that its efficiency
can exceed one hundred percent
\cite{Dadhich:1986,Dadhich:2018gmh}.} \cite{Wald:1974kya}.

It is worth noting that the joint idea in all of these mechanisms
for extracting energy comes from the PP i.e., particle fission. In
other words, this process involves the fragmentation of particles,
where the energy and angular momentum are redistributed such that
the negative energy part is absorbed by the black hole, decreasing
its mass accordingly.  Among the energy extraction mechanisms mentioned above, the BZP is the leading one which is based on the magnetic field tension around black holes.
More precisely, in the BZP, the extraction of energy occurs because the field lines counteract the spin of the black hole, and the energy is expelled by the jets that result from the winding of the magnetic field lines in the polar region \footnote{It is worth
recalling that before the BZP,  another model based on the
magnetic field for energy extraction was proposed by Ruffini and
Wilson \cite{Ruffini:1975ne}, which was not considered feasible
due to very low efficiency.}. There are phenomenologically cases
that lead to energy extraction scenarios in which the magnetic
field plays a key role. Especially since the EHT collaboration has
been able to record the magnetic field around the supermassive
black hole M87* \cite{EventHorizonTelescope:2021bee}. Furthermore,
observations of the broad Fe K$_{\alpha}$ line in the bright
Seyfert galaxy 1 MCG-6-30-15 strongly suggests the extraction of
rotational energy from the Kerr black hole through the magnetic
field \cite{Wilms:2001cm}. This idea also there is that the
magnetized plasma around a black hole can capture the trace of
particle dark matter via axion--photon interaction
\cite{Khodadi:2022ulo}.

Recently, Comisso and Asenjo \cite{Comisso:2020ykg} proposed a
novel energy extraction process, called Comisso--Asenjo process
(CAP), which is based on the magnetic reconnection occurring in
the ergosphere of a rapidly rotating black hole. This process
provides compelling evidence that the ergosphere of a black hole
is a promising source for extracting energy via magnetic
reconnection. In contrast to the PP, which relies on mechanical
particle fission to redistribute angular momentum and results in
low energy extraction rates, the CAP involves large--scale bulk
plasma motions governed by the reconnection of magnetic fields to
extract energy much more efficiently from the ergosphere of a
rapidly rotating black hole. Through the reconnection of the
magnetic field lines, a pair of plasma outflows with equal
velocities in opposite directions are generated. One of these
plasma outflows is driven against the black hole's spin, while the
other is driven in the direction of the spin, allowing it to
escape the black hole's gravitational pull and release power
provided that the plasma swallowed by the black hole has negative
energy. Magnetic reconnection is a well--known process in our
solar system, where it is commonly observed to convert magnetic
field energy into plasma kinetic energy and other forms
\cite{Jafari:2018giq}.

While black holes are commonly predicted by general relativity,
alternative theories of gravity also address their existence as
astrophysical objects. There are strong theoretical and
phenomenological motivations to support the idea that Einstein's
gravity is just an effective theory that needs to be supplemented
by a more general one (see the review papers
\cite{Clifton:2011jh,Capozziello:2011et} for more). This implies
that Kerr's black hole is an effective solution, and providing
templates based on different gravitational theories seems to be
unavoidable. Although strong gravity regime tests such as the
shadow of the supermassive black holes M87* and Sgr A*
\cite{EventHorizonTelescope:2019dse,EventHorizonTelescope:2022wkp}
and the gravitational waves from the merger of two black holes
\cite{LIGOScientific:2016aoc}  provide support for the Kerr black
hole predicted by Einstein's theory of gravity, some modified
versions still have a chance to compete for survival. Especially,
the analyzes released after the former, explicitly indicate that
some possible alternatives for the standard black holes still can
provide us with a compatible description of the object observed in
the sky (e.g., Refs.
\cite{Zakharov:2019jio,Konoplya:2019fpy,Kumar:2020hgm,Khodabakhshi:2020hny,Khodadi:2020gns,Konoplya:2021slg}).
In other words, great caution is needed when interpreting black
hole images to test Einstein's gravity \cite{Mizuno:2018lxz}. 
In
this regard, one of the interesting alternatives for the
Einstein's gravity that is of interest to us in this paper was put
forward by Moffat \cite{Moffat:2005si}. By adding a massive vector
field to the action of Einstein's gravity together with
substituting scalar fields instead of constants of the ordinary
theory, Moffat made scalar--tensor--vector gravity which is named
MOdified gravity (MOG) in the literature. 
This covariant extended
framework of gravity enjoys several worthy features such as the
successful explanation of Solar system observations
\cite{Moffat:2005si,Chan:2022ixq}, cosmological observations and data 
\cite{Moffat:2011rp,Moffat:2015bda,Negrelli:2020bkz}, dynamics of galaxy clusters
\cite{Moffat:2013uaa,Haghighi:2016jfw,DeMartino:2017ztt,Moffat:2020nmq,Harikumar:2022rxq},
and galactic rotational curves
\cite{Brownstein:2005zz,Moffat:2013sja,Moffat:2014pia}.
Importantly, MOG spares us from searching for the unknown particle
attributed to dark matter and addresses it naturally
\footnote{This is not exclusive to MOG, and there is another
alternative for Einstein's gravity so-called Modified Newtonian
Dynamics theory (MOND) \cite{milg}, which makes such a claim, too.
} \cite{Moffat:2007ju,Davari:2021mge}. 
Concerning the MOG static
and rotating black hole solutions, Moffat himself derived both
cases in \cite{Moffat:2014aja} containing an additional parameter
relative to Kerr. 
The subsequent works have investigated their
applications by serving different scenarios (e.g., see Refs.
\cite{Mureika:2015sda,Moffat:2016gkd,LopezArmengol:2016nyi,Lee:2017fbq,Perez:2017spz,Sharif:2017owq,Sheoran:2017dwb,Haydarov:2020xnv,Qiao:2020fta,Qin:2022kaf,Sheikhahmadi:2023jpb}).
To aim to get an intuition of the phenomenological imprints of MOG
black holes in strong gravity regime--based tests various studies
were conducted on gravitational lensing
\cite{Moffat:2008gi,Moffat:2014asa,Ovgun:2018fte,Rahvar:2018nhx,Izmailov:2019uhy}
and supermassive black hole shadow
\cite{Moffat:2015kva,Guo:2018kis,Moffat:2019uxp,Vagnozzi:2022moj,Kuang:2022ojj,Hu:2022lek,Hu:2023bzy}. These studies have resulted in some restrictions for additional parameter
induced by MOG. 
It was also tested by other gravitational sources such as Neutron stars \cite{LopezArmengol:2016irf} and globular clusters in the Milky Way \cite{Moffat:2007yg}.

Regarding these constraints, especially those obtained from the
shadows recorded from M87*, and Sgr A* by the EHT for MOG spacetime,
we would like to investigate the CAP--based energy extraction from
the supermassive black holes in Kerr--MOG spacetime. Namely, the
black hole that hosts the CAP is assumed to be in Kerr--MOG
spacetime, where the value of the free parameter is constrained by
observations of supermassive black holes at the centers of the
galaxies M87* and Milk Way.  
This makes our study of energy
extraction from the Kerr--MOG black hole more realistic. Other
studies on the energy extraction by Comisso--Asenjo process
\cite{Comisso:2020ykg} in other gravity theories have been carried
out in
\cite{Khodadi:2022dff,Wei:2022jbi,Liu:2022qnr,Carleo:2022qlv,Wang:2022qmg,Li:2023nmy,Li:2023htz}. Apart from rotating black holes, horizonless compact objects such as spinning wormholes are also prone to host CAP \cite{Ye:2023xyv}.

Our aim in this work is to explore whether the geometrical
modification appearing in Kerr-MOG spacetimes works in favor of
the energy extraction by CAP or not. In
connection with the spacetime at hand, among the literature we
find two ones that by employing superradiant scattering
\cite{Wondrak:2018fza}, and PP \cite{Pradhan:2018usf}
have addressed the energy extraction topic in the framework of
Kerr--MOG black hole. Interestingly, in both mechanisms,
MOG by leaving the imprints distinguishable from standard Kerr
results in diminished energy extraction. In contrast to these two
ones \footnote{Very recent by investigating analogue of gravitational synchrotron massive particle and PP in \cite{Turimov:2023mia} it is shown that Kerr--MOG black hole has an efficent role in extracting energy from black hole.}, our promising results in this paper reveal that Kerr--MOG
black hole can be a more efficient host relative to Kerr for
CAP--based energy harvesting. Throughout this paper we choose the
unit $G_N=c=1$.

The outline of this paper is as follows. In Section \ref{Ge},
after an overview of the Kerr--MOG black hole solution, we
investigate the geodesics of a particle around it and subsequently
derive some characteristics that are closely connected to the next
analysis. In Section \ref{EE}, by focusing on the role of the
deformation parameter induced by MOG and exploring the possible
parameter regions, we compute the energy extraction from the black
hole via the CAP. In Section \ref{EF}, we provide an analysis of
the power and efficiency. Finally, we release a summary of results
acquired in this paper in Section \ref{Co}.
\section{Kerr--MOG black hole and geodesics of particles around it}\label{Ge}
While the standard Kerr metric is fully characterized by mass $M$
and spin $a$,  the Kerr--MOG black hole metric \footnote{In
deriving this metric, Moffat employed two assumptions. First, due
to the smallness of the mass of the Proca--like vector field added
in the Einstein--Hilbert action, ignored its mass. Second, the
gravitational coupling $G$ is a universal constant and independent
of spacetime i.e., $\partial_{\mu}G=0$. }  introduces an
additional positive deformation parameter $\alpha$ (also known as
the MOG parameter) in the line element expressed in
Boyer--Lindquist coordinates $(t,r,\theta,\phi)$.  The line element
is given by \cite{Moffat:2014aja}
\begin{eqnarray}
    \label{metric}
    ds^2 &=& -\left(\frac{-a^2\sin^{2}\theta+\Delta}{\Sigma}\right) dt^2
    +\frac{\Sigma}{\Delta}dr^{2}-2\left(\frac{r^2+a^2-\Delta}{\Sigma}\right)a \sin^{2}\theta dt d\phi
    +\Sigma d\theta^2\nonumber\\
    &+&\left[\frac{-\Delta a^{2}\sin^2\theta +\left(r^2+a^2\right)^2}{\Sigma}\right]\sin^2\theta d\phi^2,
\end{eqnarray}
where
\begin{eqnarray}\label{del}
    \Delta&=&r^2-2GMr+a^2+M^2\alpha\, G\, G_N,\nonumber\\
    \Sigma&=&r^2+a^2\cos^2\theta .
\end{eqnarray}
According to Mofat's version \cite{Moffat:2014aja}, here $G$
denotes an enhanced gravitational constant which with the aid of
Newton's gravitational constant $G_{N}$ is parametrized in terms
of MOG parameter in the form $G=G_{N}(1+\alpha)$.  However, by
computing the ADM mass of the metric (\ref{metric}) in
\cite{Sheoran:2017dwb}, it was found that factor $1+\alpha$ indeed
affects the mass in the form $\mathcal{M}=M(1+\alpha)$, meaning
that $G=G_N$.  We have to emphasize that since according to the
Hamiltonian formalism of Einstein's gravity \footnote{In the
Einstein equations the parameter $M$ is merely an integration
constant and is not the value of the Hamiltonian for the metric
(\ref{metric}), meaning that it can not be interpreted as the
black hole mass \cite{Sheoran:2017dwb}.}, the ADM mass corresponds
to the value of the energy, thereby, here the ADM mass
$\mathcal{M}$ is, that plays the role of the gravitational mass of
the Kerr--MOG black hole. Namely, the gravitational force felt by
a test particle like a star, moving around the Kerr--MOG black
hole is essentially generated by $\mathcal{M}$.  As a result,
$\Delta$  in (\ref{del}) can be re-expressed as follows
\begin{equation}
    \label{Deltaa}
    \Delta=r^2-2\mathcal{M} r + a^2 + \frac{\alpha}{1+\alpha}\mathcal{M}^2.
\end{equation}
Due to the key role of the ergosphere in this study, deriving the
stationary limit surface (SLS), and the event horizon, for
the Kerr--MOG black hole at hand is essential. So, using
$g_{tt}=0$, and $g_{rr}=0$, for the equatorial plane
$\theta=\pi/2$, we have
\begin{eqnarray}
\label{sls1}
&&r_{\text{SLS}}= \mathcal{M} + \frac{\mathcal{M}}{\sqrt{1+ \alpha}},\\
\label{EH1}
&&r_{\text{EH}}=\mathcal{M} + \sqrt{\frac{\mathcal{M}^2}{1+ \alpha}-a^2},
\end{eqnarray}
where $r_{\text{SLS}}$ and $r_{\text{EH}}$ are the outer SLS and outer event
horizon, respectively. The ergosphere region indeed is in the
range from $r_{\text{SLS}}$ to $r_{\text{EH}}$. Note that the positiveness of
the radical in Eq. (\ref{EH1}) impose the following bounds on the
MOG and spin parameters
\begin{eqnarray}
    \label{alphabounds1}
&&  0\le\alpha\le \frac{\mathcal{M}^2}{a^2}-1,
\end{eqnarray}
meaning that it ensures we deal with a black hole configuration, instead of a naked singularity which is free of the event horizon.
For convenience, in some situations, we use the dimensionless parameter $\tilde{a}=a/\mathcal{M}$.

Following the procedure outlined in the seminal reference \cite{Bardeen:1972fi}, the geodesics for particles in the Kerr--MOG black hole background take the following form
\begin{eqnarray}
    \Delta\Sigma\,\frac{dt}{d\lambda}\!=\!\left[(r^2\!+\!a^2)^2\!-\!\Delta a^2 \sin^2\!\theta \right]\!E\!-\!(r^2\!+\!a^2\!-\!\Delta)aL,   \label{Ut}
\end{eqnarray}
\begin{eqnarray}
    \Sigma^2 (\frac{dr}{d\lambda})^2\!=\!T^2\!-\!\Delta\left[r^2\!+\!(L\!-\!aE)^2\!+\!{\cal K}\right] \equiv V_r^2(r),
    \label{V2}
\end{eqnarray}
\begin{equation}
    \Sigma^2(\frac{d\theta}{d\lambda})^2\!=\!{\cal K}\!-\!\left[a^2(1\!-\!E^2)\!+\!\frac{L^2}{\sin^2 \theta}\right]\cos^2\theta\!\equiv\!\Theta^2(\theta),
    \label{Theta2}
\end{equation}
\begin{eqnarray}
    \!\Delta\Sigma\sin^2\!\theta \frac{d\phi}{d\lambda}\!\!=\!a\sin^2\!\theta(r^2\!+\!a^2\!\!-\!\Delta\!)E\!+
    \!(\Delta\!-\!a^2\sin^2\!\theta)L,  \label{Uphi}
\end{eqnarray}
where
\begin{eqnarray}
    \label{T}
    T\equiv E\left(r^2+a^2\right)-La.
\end{eqnarray} Here $V_r^2(r)$ and $\Theta^2(\theta)$ respectively are $r$ and $\theta$ dependent functions. $E$, $L$, ${\cal K}$ also are the constants of motion and $\lambda$ denotes the affine parameter along the geodesics. By adopting the assumption that magnetic reconnection happens in the bulk plasma that circularly rotates around the black hole at the equatorial plane
$\theta=\pi/2$, one can set $\frac{d\theta}{d\lambda}=0$, and subsequently ${\cal K}=0$ since $\Theta^2(\theta)$ vanishes.
Now, Eqs. (\ref{Ut})--(\ref{Uphi}), take the following simple forms
\begin{eqnarray}
    \label{tdot}
    r^2\frac{dt}{d\lambda}&=&a\left(L-aE\right)+\left(r^2+a^2\right)\frac{T}{\Delta},\\
    \label{rdot}
    r^2\frac{dr}{d\lambda}&=&\pm \sqrt{V_{r}},\\
    \label{pdot}
    r^2\frac{d\phi}{d\lambda}&=& \left(L-aE\right)+\frac{aT}{\Delta}.
\end{eqnarray}
Here $V_{r}$ indeed is an ``effective potential" governing the motion of particle in the radial coordinate $r$. To have bounded circular orbits for particles in the equatorial plane it is essential that the radial component in (\ref{rdot}) vanishes at a fixed distance $r$ (i.e., $\frac{dr}{d\lambda}=0$), and possess a minimum, as well. Thus, Eq. (\ref{rdot}) gives the following condition on the effective potential as well as on its first derivative \cite{Bardeen:1972fi,Wilkins:1972rs}
\begin{equation}
    \label{Vmin}
    V_{r}=0,   \qquad \mbox{ and} \qquad \frac{dV_{r}}{dr}=0.
\end{equation}
After solving these equations for the energy $E$, and the angular momentum $L$, as conserved quantities, respectively we come to the following expressions  \footnote{To find further details of the calculation of these two conserved quantities refers to the appendix of Ref. \cite{Sheoran:2017dwb}.}
\begin{eqnarray}
    \label{E1}
    E&=&\frac{1}{\sqrt{Q_\pm}}\left(1-\frac{2\mathcal{M}}{r}+
    \mathcal{A}\frac{\mathcal{M}}{r^2}\pm a \sqrt{\frac{\varpi}{r^3}}\right),\\
    \label{L1}
    L&=& \pm \sqrt{\frac{r\varpi}{Q_\pm}}\left(\frac{a^2}{r^2}+1\mp a\sqrt{\frac{1}{r^3\varpi}}\left(2\mathcal{M}-\mathcal{A}\frac{\mathcal{M}}{r}\right)\right),
\end{eqnarray}
where
\begin{eqnarray}
    \mathcal{A}&=&\frac{\alpha}{1+\alpha},\\
    Q_\pm&=&1+2\mathcal{A}\frac{\mathcal{M}^2}{r^2}-\frac{3\mathcal{M}}{r}\pm 2a\sqrt{\frac{\varpi}{r^3}},\\
    \varpi&=& \mathcal{M}-\mathcal{A}\frac{\mathcal{M}}{r}. \label{pi}
\end{eqnarray}
By plugging the expressions above into Eqs. (\ref{tdot}), and (\ref{pdot}), we obtain the following relevant expressions for
Keplerian angular velocity i.e., $\Omega_K= \frac{U^{\phi}}{U^{t}}$
\begin{eqnarray}
    \label{tdot1}
    U^t&=&\frac{r^{\frac{3}{2}}\pm a\sqrt{\varpi}}{r^{\frac{3}{2}}\sqrt{Q_{\pm}}},\\
    \label{pdot1}
    U^{\phi}&=&\frac{\pm \sqrt{\varpi}}{r^{\frac{3}{2}}\sqrt{Q_{\pm}}}.
\end{eqnarray}
It is easy to check that by relaxing the MOG parameter ($\alpha\rightarrow 0$), the Keplerian angular velocity of Kerr black hole i.e., $\Omega_K=\frac{\pm\sqrt{M}}{r^{3/2}\pm a\sqrt{M}}$, recovers. Signs +, and - respectively denote co-rotating and counter--rotating orbits. The former assures us that magnetic reconnection happens inside the ergosphere, so we take care of it throughout the paper.

Another quantity that will be needed later is the circular photon orbit $r_{\text{ph}}$. For the case of non-rotating black hole, it is written as follows \cite{Moffat:2014aja}
\begin{equation}
\label{rp}
r_{\text{ph}}=\frac{(1+\alpha)M}{2}\left(3+\sqrt{\frac{9+\alpha}{\alpha+1}}\right),
\end{equation} where by taking the definition of ADM mass into it, reads as
\begin{equation}
\label{rp1}
r_{\text{ph}}=\frac{\mathcal{M}}{2}\left(3+\sqrt{\frac{9+\alpha}{\alpha+1}}\right).
\end{equation}At first eye, one sees that by relaxing $\alpha$, it is equal to $r_{\text{ph}}=3M$, as expected for standard Schwarzschild. Concerning the Kerr--MOG black hole's innermost stable circular orbit (ISCO), a comprehensive analysis has already been done in \cite{Lee:2017fbq}, which gives a long--expression for the circular orbit of the photon. Given that we are interested in co--rotating (prograde) orbits around a rapidly rotating black hole, one concludes from \cite{Lee:2017fbq} that for small values of MOG parameter $\alpha$, and $\tilde{a}\sim1$, then $r_{\text{ph}}$ for the Ker--MOG black hole mets standard Kerr i.e., $r_{\text{ph}}\sim \mathcal{M}$.
As a result, we will set $r_{\text{ph}}\sim \mathcal{M}$ for the prograde photon circular orbit of the Ker--MOG black hole at hand.


\section{Energy extraction via CAP }\label{EE}
In what follows, inspired by Ref. \cite{Comisso:2020ykg}, we first present the main body of the CAP that occurs in the ergosphere of a black hole as the generator of energy extraction. Then, we utilize it within the Kerr--MOG black hole as a supermassive host for the CAP.  To guarantee this, we need to adjust the MOG parameter according to the constraints obtained from the EHT observations of the supermassive black holes M87*, and Sgr A* \cite{Moffat:2015kva,Guo:2018kis,Moffat:2019uxp,Vagnozzi:2022moj,Kuang:2022ojj,Hu:2022lek,Hu:2023bzy}. Among these, the most tight constraint on the MOG parameter released in Ref. \cite{Vagnozzi:2022moj} i.e., $\alpha \lesssim 0.01$.
Based on bound (\ref{alphabounds1}), guaranteeing a black hole configuration, this constraint indicates that the value of the rotation parameter of the MOG black hole is near--extremal i.e., $\tilde{a}\sim0.99$.
Throughout this manuscript, we use values of the MOG parameter that satisfy this upper bound.  In this way, one really can be imagined that the host black hole suggested by MOG spacetime addresses a supermassive one.
\subsection{Outline of the CAP}
The formation of flux ropes/plasmoids \cite{Comisso:2016pyg} in the equatorial plane of a black hole can lead to fast magnetic reconnection if the current sheet exceeds a critical aspect ratio \cite{Comisso:2016pyg,Uzdensky:2014uda,Comisso:2017arh}. Due to fast--happening magnetic reconnection then the accessible magnetic energy is released in the form of plasma particles that are expelled from the reconnection layer \cite{Jafari:2018giq}. More precisely, based on the idea supported by some numerical simulations \cite{Parfrey:2018dnc,Ripperda:2020bpz,Bransgrove:2021heo}, the reconnection of magnetic field lines takes place at a dominant X--point and subsequently the plasma containing flux ropes/plasmoids is ejected. So, the dominant X--point is, in essence, the location of magnetic reconnection.  In the ergosphere of a black hole,  the plasmas moving in opposite directions experience negative and positive accelerations. A distant observer would see the plasma with negative acceleration gaining negative energy and eventually being swallowed by the black hole. On the other hand, the plasma with positive acceleration carry more energy due to conservation of energy, leading to the energy extraction from the black hole.
In the process of extracting energy from a black hole, it is crucial to note that the decelerated plasma must have negative energy from the perspective of a distant observer. This phenomenon can only occur if the X--point is located within the ergosphere of the black hole, i.e., a region between the event horizon and SLS.

Before going into further details,  it is important to note the significant differences between the magnetic reconnection process proposed by Comisso and Asenjo and other magnetic reconnection models, such as the one proposed by Koide and Arai in Ref. \cite{Koide:2008xr} or Gouveia Dal Pino and collaborators in Refs. \cite{deGouveiaDalPino:2003mu,Kadowaki:2014nda,Singh:2014jma,deGouveiaDalPino:2016tsr}. In the CAP, the reconnection zone is filled with flux ropes/plasmoids, which gives rise to fast magnetic reconnection while in \cite{Koide:2008xr} the reconnection zone is laminar,  which gives rise to slow magnetic reconnection. For Refs. \cite{deGouveiaDalPino:2003mu,Kadowaki:2014nda,Singh:2014jma,deGouveiaDalPino:2016tsr}, the reconnection is in the collisional turbulent regime,  which typically gives a fast reconnection regime.  
The differences in the reconnection zones lead to different inflow and outflow velocity formulas between \cite{Comisso:2020ykg},  \cite{Koide:2008xr},  and \cite{deGouveiaDalPino:2003mu,Kadowaki:2014nda,Singh:2014jma,deGouveiaDalPino:2016tsr}.  The effectiveness of the magnetic reconnection process is therefore strongly regulated by these differences.  For a magnetic reconnection process that is slow,  the extremely low magnetic diffusion that characterizes the accretion flows leads to an almost negligible reconnection process, resulting in little to no energy extraction.  Another important difference concerns the inflow and outflow velocities in these magnetic reconnection models,  and this difference is essential, as shown in \cite{Comisso:2020ykg}.  Specifically, in the case of \cite{Comisso:2020ykg} the electric current sheet of the reconnection layer lies in the equatorial plane, whereas the inflow velocity is in the meridional plane, and the outflow velocity is in the equatorial plane.  
In \cite{Koide:2008xr} the electric current sheet of the reconnection layer is in the plane perpendicular to the equatorial plane,  while both inflow and outflow velocities are in the equatorial plane.  
This important difference prevents net power extraction and any evaluation of the efficiency in \cite{Koide:2008xr}.  On the other hand,  in the models discussed in Refs. \cite{deGouveiaDalPino:2003mu,Kadowaki:2014nda,Singh:2014jma,deGouveiaDalPino:2016tsr}, the magnetic reconnection occurs between the corona of the accretion disk and the jet of the black hole, meaning that the X--point is located outside the ergosphere,  and no energy can be extracted from the black hole.

Now let us address analytically the plasma energy density which is essential for energy extraction. To do this, a \textit{``zero angular momentum observer''} (ZAMO) frame is usually used, which is a non-rotating local frame, with the following line element
\begin{eqnarray}
    ds^2=-d\hat{t}^2+\sum_{i=1}^3(d\hat{x}^{i})^2=\eta_{\mu\nu}dx^{\mu}dx^{\nu},
\end{eqnarray}
where \footnote{Note that throughout this paper $\boldsymbol{\alpha}$ is different from $\alpha$ and indeed they address two different ones.}
\begin{eqnarray}
    d\hat{t}=\boldsymbol{\alpha} dt,\quad d\hat{x}^{i}=\sqrt{g_{ii}}dx^{i}-\boldsymbol{\alpha}\beta^{i}dt.
\end{eqnarray}
with
\begin{eqnarray}
    \label{alpha}
\boldsymbol{\alpha}&=&\sqrt{-g_{tt}+\frac{g_{\phi t}^2}{g_{\phi\phi}}},\\
\label{alpha1}
\beta^{i}&=&(0, 0, \beta^{\phi})=(0,0,  \frac{\sqrt{g_{\phi\phi}}\omega^{\phi}}
{\boldsymbol{\alpha}}).
\end{eqnarray}
Here $\boldsymbol{\alpha}$,  $\beta^{i}$, and $\omega^{\phi}$ respectively denote the lapse function, shift vector, and the angular velocity of the frame dragging which is defined as $\omega^{\phi}=-g_{\phi t}/g_{\phi\phi}$. It is required to comment that the quantities labeled with hats are ones observed in the ZAMO frame such that
one can connected them to their counterpart in Boyer-Lindquist coordinates via following transformations
\begin{eqnarray}
\hat{\psi}^0&=&\boldsymbol{\alpha} \psi^0,\quad \hat{\psi}^{i}=\sqrt{g_{ii}}\psi^{i}-\boldsymbol{\alpha}\beta^{i}\psi^0,\label{vvect}\\
\hat{\psi}_0&=&\frac{\psi_0}{\boldsymbol{\alpha}}+\sum_{i=1}^3\frac{\beta^i}{\sqrt{g_{ii}}}\psi_i,\quad
\hat{\psi}_i=\frac{\psi_i}{\sqrt{g_{ii}}},
\end{eqnarray}
for the contravariant and covariant components, respectively.

By evaluating the conditions imposed due to magnetic reconnection in the ergosphere, namely, the formation of negative energy at infinity along with the absorption of the decelerated plasma by the black hole,  and the expulsion of the accelerated plasma to infinity, we here turn to examine the capability of the CAP to harvesting energy from the black hole. By taking one-fluid approximation into account, and without addressing the origin of properties of plasma, its energy-momentum tensor is written as
\begin{eqnarray}
T^{\mu\nu}=pg^{\mu\nu}+wU^{\mu}U^{\nu}+F^{\mu}_{\;\;\;\delta}F^{\nu\delta}
-\frac{1}{4}g^{\mu\nu}F^{\rho\delta}F_{\rho\delta},
\end{eqnarray}
with $p$, $w$, $U^{\mu}$, and $F^{\mu\nu}$ which refer to
the proper plasma pressure, enthalpy density, four-velocity, and electromagnetic field tensor, respectively. According to definition $e^\infty=-\boldsymbol{\alpha} g_{\mu 0}T^{\mu 0}$ for the energy-at-infinity, we have
\begin{eqnarray}
e^\infty=\boldsymbol{\alpha} \hat{e}+\boldsymbol{\alpha} \beta^{\phi}\hat{P}^\phi,
\end{eqnarray}
where $\hat{e}$, and $\hat{P}^{\phi}$ are respectively the total energy density, and the azimuthal component of the momentum density and read as
\begin{eqnarray}
\hat{e}&=&w\hat{\gamma}^2-p+\frac{\hat{B}^2+\hat{E}^2}{2},\\
\hat{P}^{\phi}&=&w\hat{\gamma}^2\hat{v}^{\phi}+(\hat{B}\times\hat{E})^{\phi}.
\end{eqnarray}
Here, $\hat{\gamma}=\sqrt{1-\sum_{i=1}^{3}(d\hat{v}^i)^2}$ ($\hat{v}^i=(\hat{v}^{r},\hat{v}^{\theta},\hat{v}^{\phi})$ are the components of the out flow velocity of plasma in the view of a ZAMO observer) is the Lorentz factor, and the
components of electric fields, and magnetic is defined as  $\hat{E}^i=\eta^{ij}\hat{F}_{j0}=\hat{F}_{i0}$, and $\hat{B}^i=\epsilon^{ijk}\hat{F}_{jk}/2$. The energy-at-infinity density $e^{\infty}$, in essence, is a summation of the
following components
\begin{eqnarray}
e^{\infty}_{\text{hyd}}&=&\boldsymbol{\alpha}(w\hat{\gamma}^2-p)+\boldsymbol{\alpha}\beta^{\phi}w\hat{\gamma}^2\hat{v}^{\phi},\\
e^{\infty}_{\text{em}}&=&\boldsymbol{\alpha}(\hat{B}^2+\hat{E}^2)/2+\boldsymbol{\alpha}\beta^{\phi}(\hat{B}\times\hat{E})_{\phi},
\end{eqnarray} where the former is hydrodynamic and the latter is electromagnetic. In case of demanding the most efficiency for the magnetic energy arising from the magnetic reconnection process i.e., converting it to the kinetic energy of the plasma, one can ignore the negligible contribution of the electromagnetic energy in comparison with the hydrodynamic energy-at-infinity \footnote{Although fully kinetic particle-in-cell simulations (see, e.g., Ref. \cite{Sironi:2016vyg}) show that magnetic energy is converted into bulk kinetic energy and the outflow follows the scaling relations presented in our work, two-fluid simulations performed in \cite{Zenitani:2009di} contains the message is that a large fraction of magnetic energy is converted into enthalpy. 
However, it should be noted that close to the black hole, the plasma is collisionless, meaning that the fully kinetic particle-in-cell solution should be the proper one to adopt. This will become a challenging situation if the output of \cite{Zenitani:2009di} is further adapted by numerical simulations.}. As a result, by serving incompressible and adiabatic approximation for the plasma, the energy-at-infinity density can be well evaluated by the following expression
\begin{eqnarray}\label{en}
e^{\infty}=e^{\infty}_{\text{hyd}}=\boldsymbol{\alpha} w\hat{\gamma}(1+\beta^{\phi}\hat{v}^{\phi})-\frac{\boldsymbol{\alpha} p}{\hat{\gamma}}.
\end{eqnarray}
Introducing the local rest frame $x'^{\mu}$=($x'^{0}$, $x'^{1}$, $x'^{2}$, $x'^{3}$) for the bulk plasma that rotates around the black hole with Keplerian angular velocity $\Omega_{\text{K}}$, is  helpful in analyzing the localized reconnection process located in the equatorial plane. The directions $x'^{1}$, $x'^{3}$
are chosen respectively to be parallel to the radial direction, and the azimuthal direction i.e., $x^{1}=r$ and $x^{3}=\phi$. By  using the transformation (\ref{vvect}), one can write the co-rotating Keplerian velocity in the ZAMO frame as follows
\begin{eqnarray}\label{K}
    \hat{v}_{\text{K}}&=&\frac{d\hat{x}^{\phi}}{d\hat{x}^{t}}=\frac{d\hat{x}^{\phi}/d\lambda}{d\hat{x}^{t}/d\lambda}
    =\frac{\sqrt{g_{\phi\phi}}dx^{\phi}/d\lambda-\boldsymbol{\alpha}\beta^{\phi}dx^{t}/d\lambda}{\boldsymbol{\alpha} dx^{t}/d\lambda}\nonumber\\
    &=&\frac{\sqrt{g_{\phi\phi}}}{\boldsymbol{\alpha}}
    \Omega_{\text{K}}-\beta^{\phi}.
\end{eqnarray}
The explicit form of $\hat{v}_{\text{K}}$ shall acquire upon putting $\Omega_{\text{K}}$ in the equation above. $\hat{v}^{\phi}$ in (\ref{en}) is the outflow velocity measured by observer located in the ZAMO frame, and reads as
\begin{eqnarray}\label{enn}
\hat{v}^{\phi}=\frac{\hat{v}_{\text{K}}\pm v_{\text{out}} cos \xi}{1\pm\hat{v}_{\text{K}}v_{\text{out}}cos \xi},~~~v_{\text{out}}=\sqrt{\frac{\sigma_0}{1+\sigma_0}},
\end{eqnarray}
where $\hat{\gamma}_{\text{K}}=1/\sqrt{1-\hat{v}_{\text{K}}^2}$
is corresponds to the Lorentz factor, $\sigma_0$ refers to the plasma magnetization upstream of the reconnection layer, and $\xi$ is the orientation angle between the lines of magnetic field, and the azimuthal direction in the equatorial plane. Now by taking (\ref{enn}) into (\ref{en}), together with this assumption that the plasma is relativistic hot \footnote{The relativistic hot assumption works reasonably well when the temperature is only marginally relativistic, i.e., comparable to the electron rest mass energy. In the magnetic reconnection process, if some of the magnetic energy is converted into enthalpy in addition to bulk kinetic energy, we expect this scenario to occur still. In any case, taking a relativistic hot or cold plasma into account modifies the adiabatic index in the equation of state from $4/3$ to $5/3$. Inserting the latter ($5/3$) does not affect the results of the analysis because it would produce only minor modifications in the numerical factor of the last term of the hydrodynamic energy at infinity per enthalpy.} with $w=p/4$, one obtains the final form of the hydrodynamic energy at infinity per plasma enthalpy $\epsilon_{\pm}^{\infty}=e^{\infty}/w$ as follows (see Ref. \cite{Comisso:2020ykg},  for more)
\begin{eqnarray}\label{ex}
\epsilon_{\pm}^{\infty}=\boldsymbol{\alpha}\hat{\gamma}_{\text{K}}\left((1+\beta^{\phi}\hat{v}_{\text{K}})\sqrt{1+\sigma_0}\pm\cos\xi(\beta^{\phi}+\hat{v}_{\text{K}})\sqrt{\sigma_0}
-\frac{\sqrt{1+\sigma_0}\mp \cos\xi \hat{v}_{\text{K}}\sqrt{\sigma_0}}{4\hat{\gamma}^{2}_{\text{K}}(1+\sigma_0-\cos^2\xi\hat{v}^{2}_{\text{K}}\sigma)}\right).
\end{eqnarray}
The signs +, and - in the expression above denote to accelerated and decelerated plasma particles which respectively escape in co-rotating and counter-rotating directions relative to the black hole's rotation. In case of satisfying the conditions $\Delta\epsilon_{+}^{\infty}>0$, and $\epsilon_{-}^{\infty}<0$, the energy extraction takes place, meaning that in the light of the magnetic reconnection process, the energy-at-infinity of the accelerated part of the plasma becomes greater than its rest mass and thermal energies while for the decelerated part it is negative.
Due to this assumption that we deal with a relativistic hot plasma with polytropic index $\Gamma=4/3$, and $w=p/4$, thereby, $\Delta\epsilon_{+}^{\infty}=\epsilon_{+}^{\infty}-\left (1-\dfrac{\Gamma}{\Gamma-1}.\dfrac{p}{w}\right)=\epsilon_{+}^{\infty}$. So, the final form of the condition of energy extraction reads as $\epsilon_{+}^{\infty}>0$, and $\epsilon_{-}^{\infty}<0$.
\begin{figure}[ht!]
\begin{tabular}{c}
\includegraphics[width=.5\linewidth]{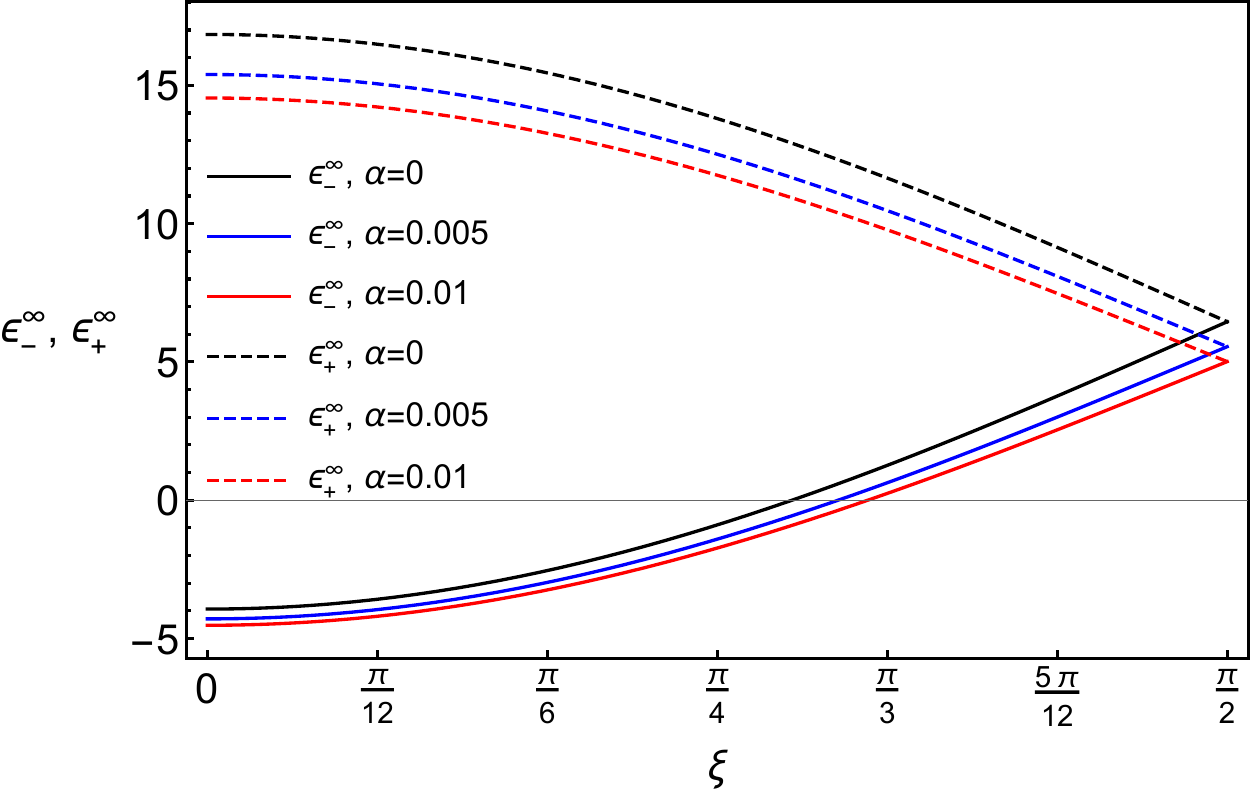}~~~
\includegraphics[width=.5\linewidth]{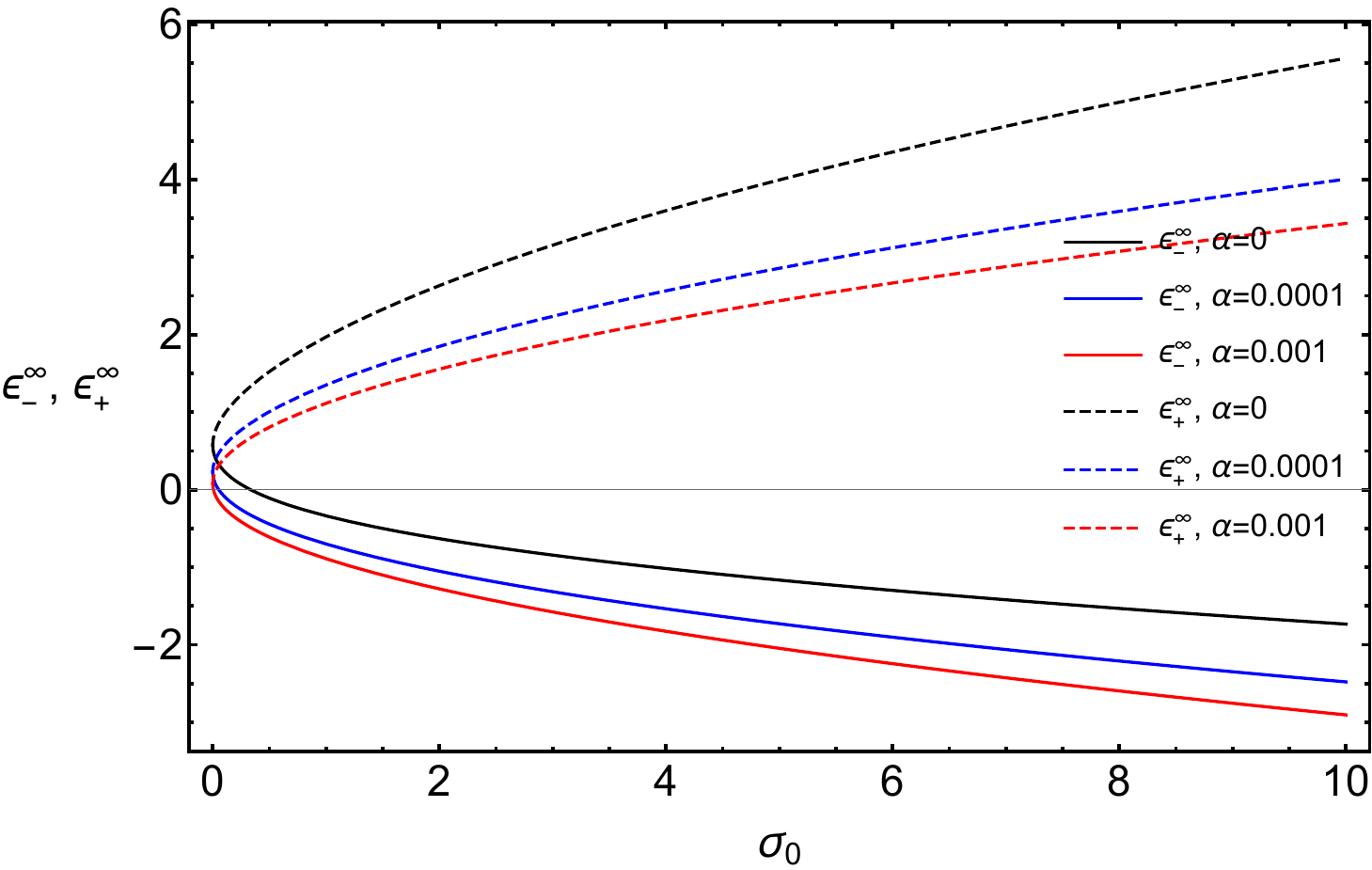}
    \end{tabular}
    \caption{$\epsilon_{\pm}$ in terms of the orientation angle $\xi$ (left panel) for near--extremal black hole
        ($\tilde{a}\sim 0.99$), with $X=1.2$, $\sigma_0=10$, and different values of MOG parameter $\alpha$. The same in terms of the plasma magnetization $\sigma_0$ with $X=1.2$, and $\xi=0$ in right panel.} \label{e1}
\end{figure}

\begin{figure}[ht!]
    \begin{tabular}{c}
        \includegraphics[width=.47\linewidth]{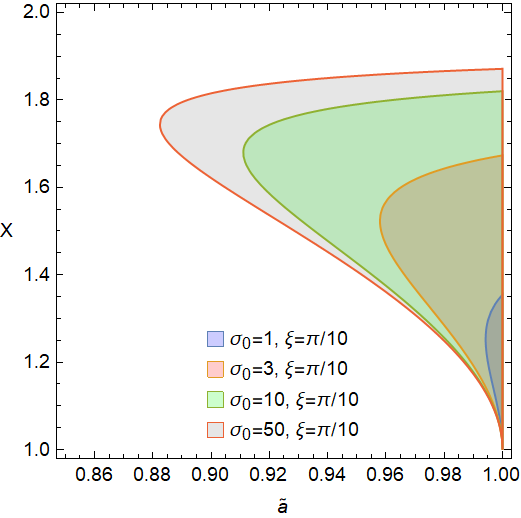}~~~
        \includegraphics[width=.47\linewidth]{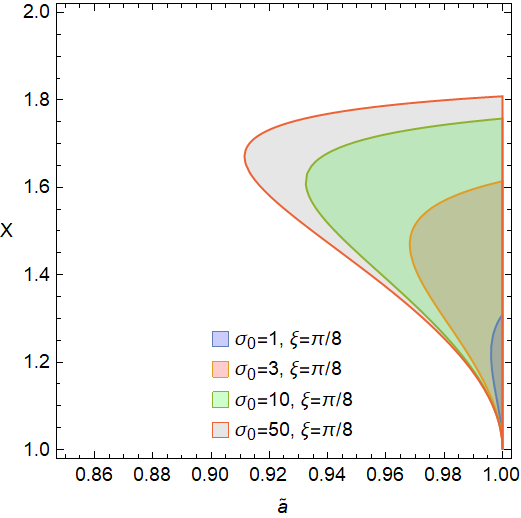}\\
        \includegraphics[width=.47\linewidth]{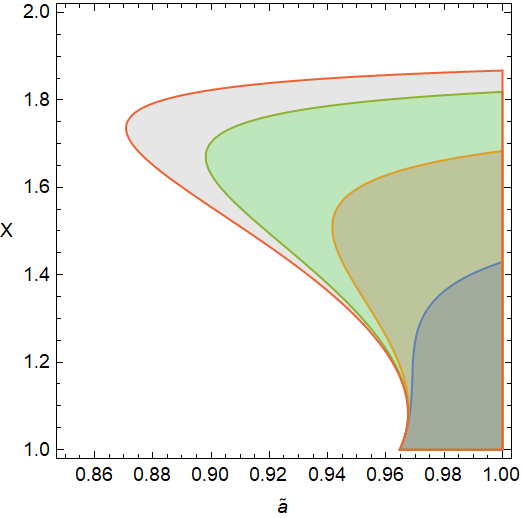}~~~
        \includegraphics[width=.47\linewidth]{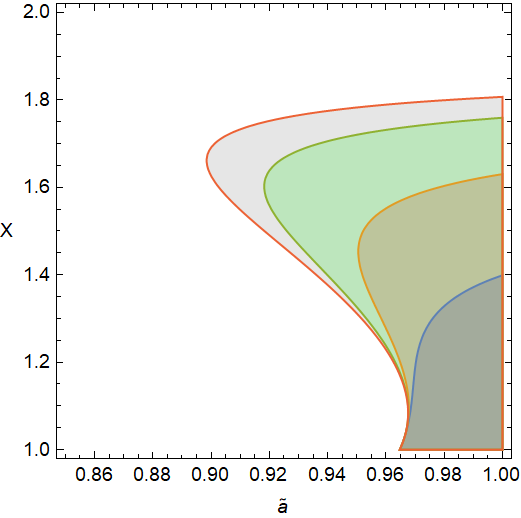}
    \end{tabular}
    \caption{ Regions of the parameter space $\tilde{a}-X$ in which energy extraction conditions $\epsilon_{-}<0$, and $\epsilon_{+}>0$ satisfy for the standard Kerr black hole (top row), and Kerr--MOG black hole with free parameter $\alpha=0.01$ (bottom row). The same numerical values used for $\sigma_0$, and $\xi$ in the top row are for the bottom, too.} \label{e2}
\end{figure}

\begin{figure}[ht!]
    \begin{tabular}{c}
        \includegraphics[width=.47\linewidth]{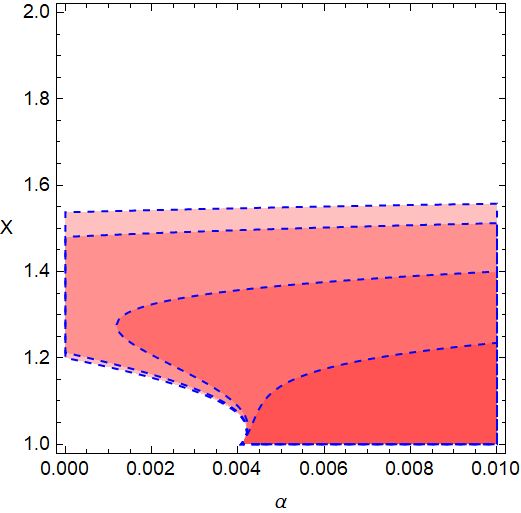}~~~
        \includegraphics[width=.47\linewidth]{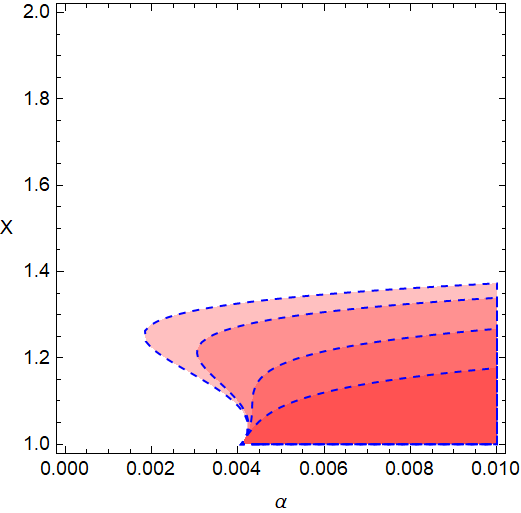}
    \end{tabular}
    \caption{Regions of the parameter space $\alpha-X$ which satisfies energy extraction conditions $\epsilon_{-}<0$, and $\epsilon_{+}>0$ for the near-extremal Kerr-MOG black hole $(\tilde{a}\sim0.99)$. The regions shaded in red (from dark to light) respectively belong to the plasma magnetization $\sigma_0\in \{1,3,10,50\}$ with orientation angle $\xi=\pi/6$ (left), and $\xi=\pi/5$ (right). Borderlines (blue-dashed) separate these areas.}
    \label{e3}
\end{figure}

\begin{figure}[ht!]
    \begin{tabular}{c}
        \includegraphics[width=.47\linewidth]{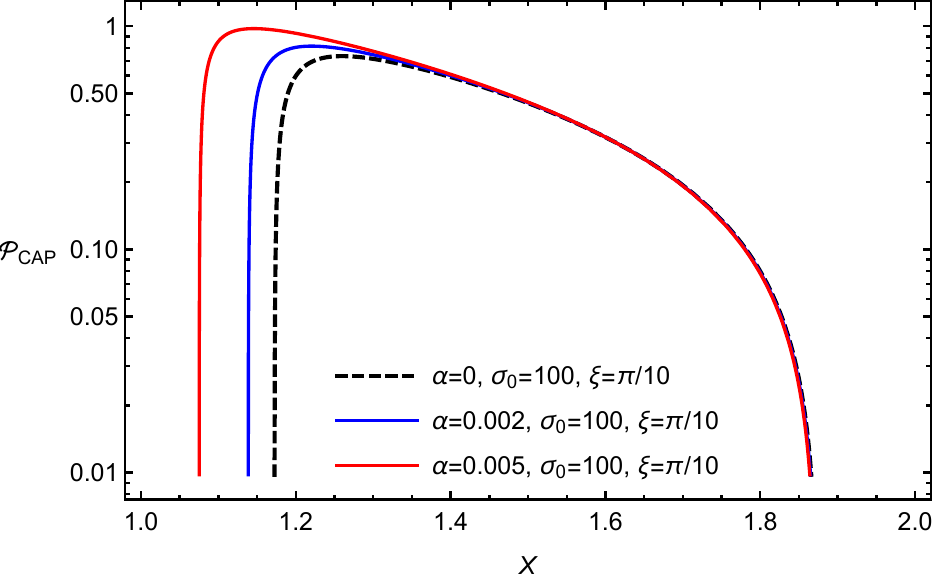}~~~
        \includegraphics[width=.47\linewidth]{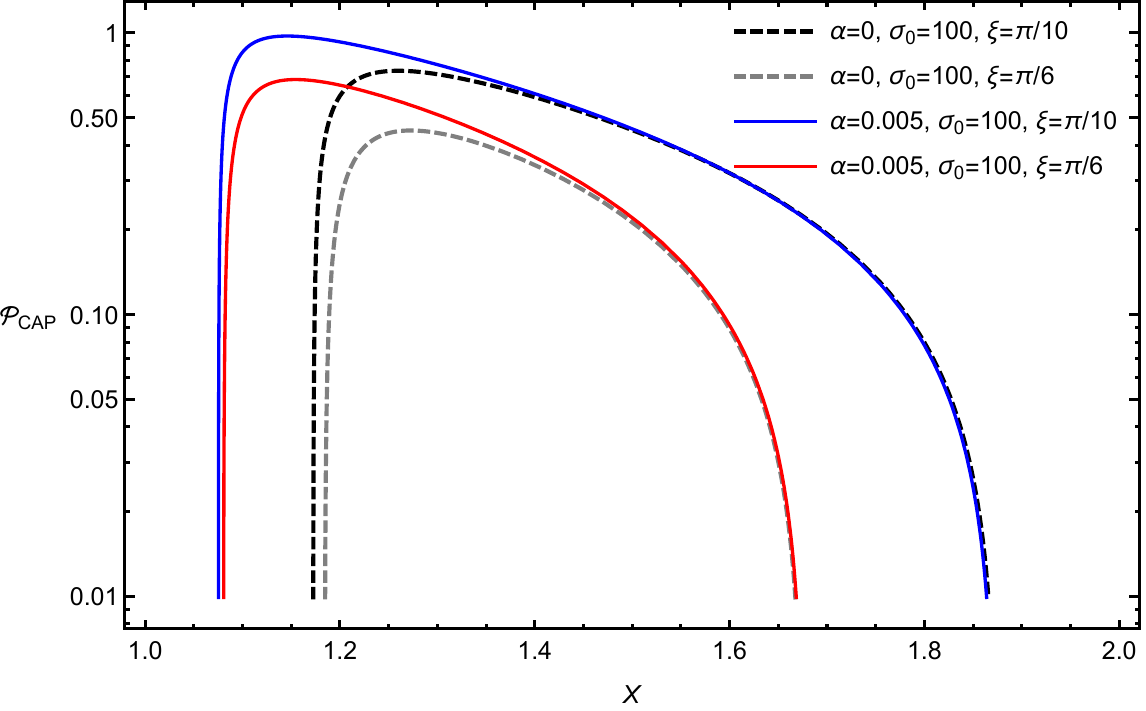}\\
        \includegraphics[width=.47\linewidth]{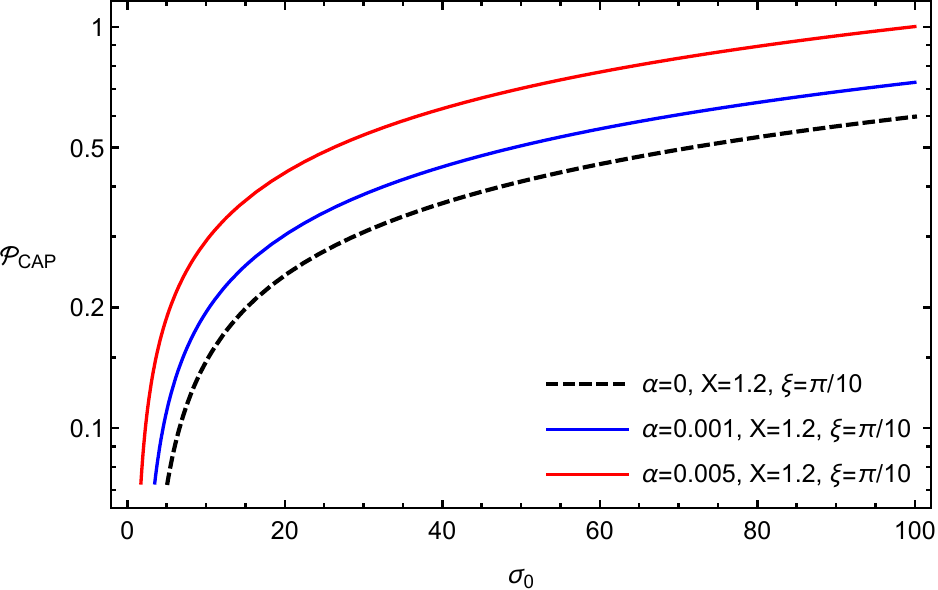}~~~
        \includegraphics[width=.47\linewidth]{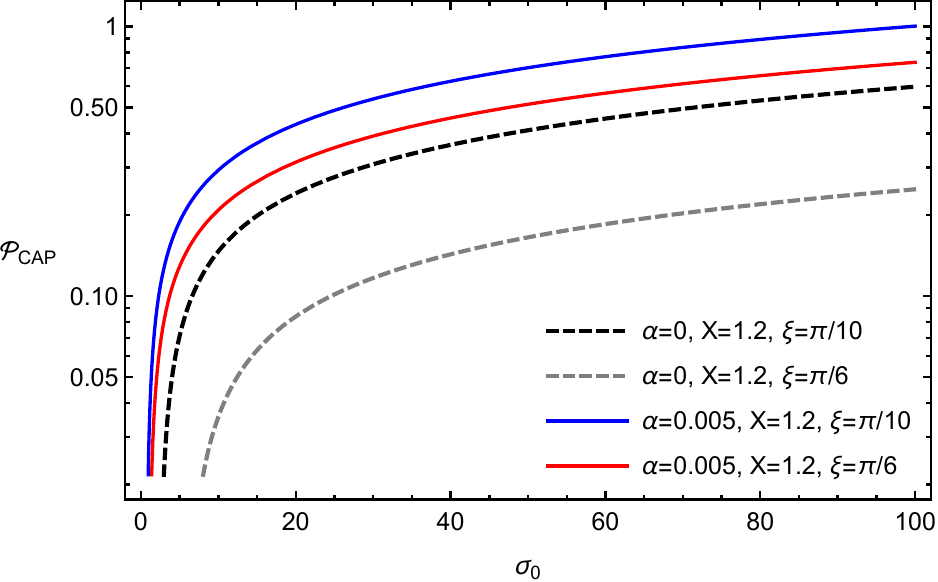}
    \end{tabular}
    \caption{$\mathcal{P}_{\text{CAP}}$ in terms of the dominant X--point location (top row) and the plasma magnetization $\sigma_0$ (bottom row), for a near--extremal black hole ($\tilde{a}\sim 0.99$), with different values of MOG parameter $\alpha$. } \label{e4}
\end{figure}

\begin{figure}[ht!]
    \begin{tabular}{c}
        \includegraphics[width=.47\linewidth]{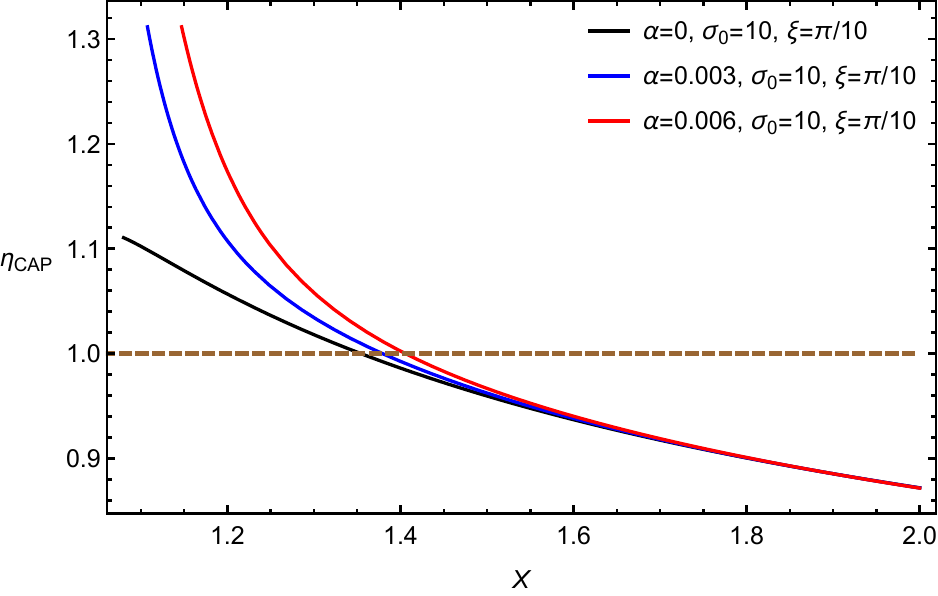}~~~
        \includegraphics[width=.47\linewidth]{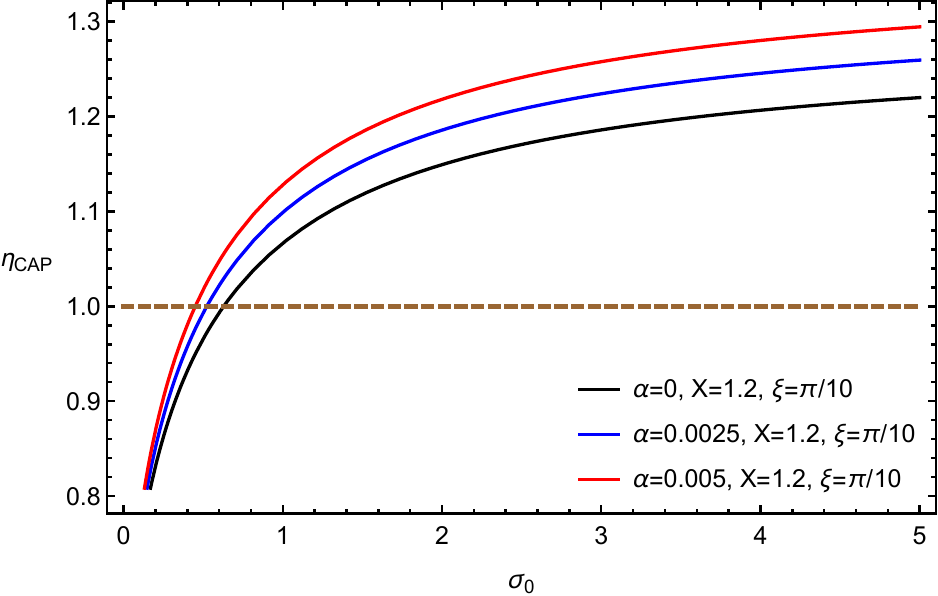}\\
        \end{tabular}
    \caption{$\eta_{\text{CAP}}$ in terms of the dominant X--point location (left), and the plasma magnetization $\sigma_0$ (right), for a near--extremal black hole ($\tilde{a}\sim 0.99$), with different values of MOG parameter $\alpha$. The brown-dashed line represents the borderline $\eta_{\text{CAP}}=1$.} \label{e5}
\end{figure}

\begin{figure}[ht!]
    \begin{tabular}{c}
        \includegraphics[width=.47\linewidth]{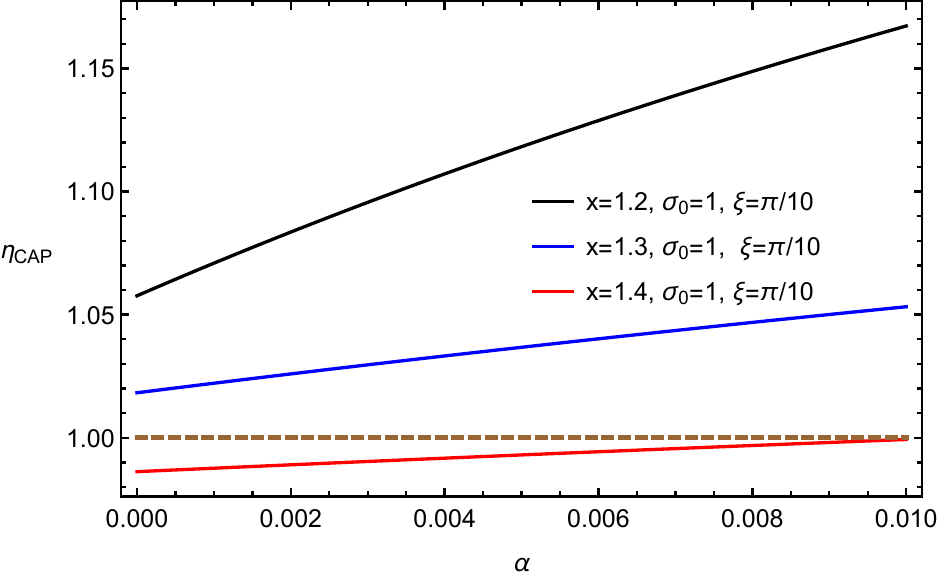}~~~
        \includegraphics[width=.47\linewidth]{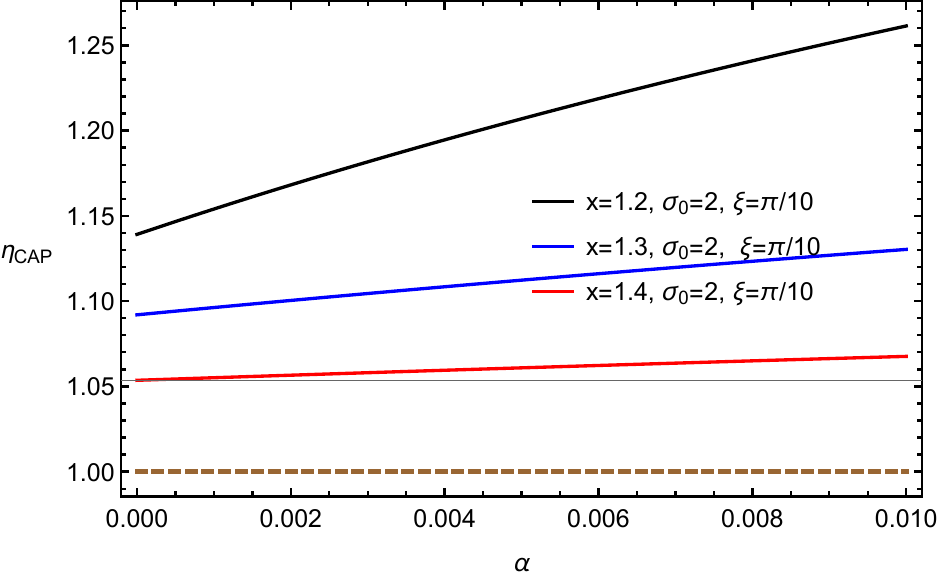}\\
    \end{tabular}
    \caption{$\eta_{\text{CAP}}$ in terms of the MOG parameter $\alpha$ for different values of the plasma magnetization $\sigma_0$ (left and right panels). } \label{e6}
\end{figure}

\begin{figure}[ht!]
        \centering
    \begin{tabular}{c}
        \includegraphics[width=.6\linewidth]{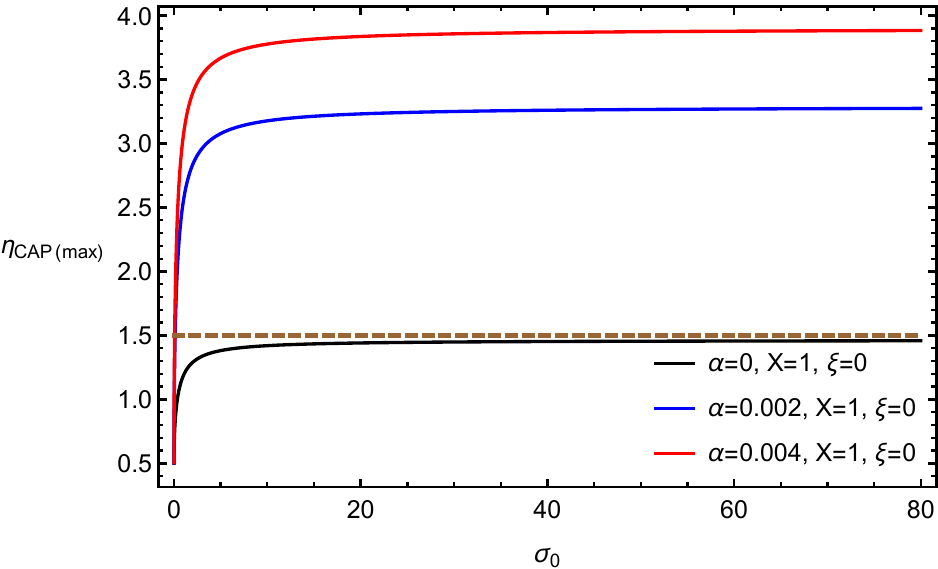}~~~
    \end{tabular}
    \caption{$\eta_{\text{CAP(max)}}$ in terms of the the plasma magnetization $\sigma_0$ for different values of MOG parameter $\alpha$. The brown-dashed line represents the borderline $\eta_{\text{CAP(max)}}=3/2$.} \label{ee6}
\end{figure}
\subsection{Energy extraction in the presence of MOG parameter}
At this stage, we can reveal the role of the MOG parameter in the CAP-based energy extraction from a rotating black hole.
However, the explicit form of $\hat{v}_{\text{K}}$ in the expression (\ref{ex}) is still unspecified. By calculating $\boldsymbol{\alpha}$ and $\beta^{\phi}$ for the components of the metric of MOG spacetime
\begin{eqnarray}
\label{}
\boldsymbol{\alpha}&=&\sqrt{\frac{\tilde{a}^2 \left(-\alpha ^2+(\alpha +1)^2 X^4+4 \alpha  (\alpha +1) X\right)+(\alpha +1) X^4 (\alpha +(\alpha +1) X(X-2))}{(\alpha +1) X^2 \left(\tilde{a}^2 ((\alpha +1) X (X+2)-\alpha )+(\alpha +1) X^4\right)}}, \\
\label{}
\beta^{\phi}&=&\frac{2 \tilde{a}}{\sqrt{\tilde{a}^2 \left((\alpha +1)^{-1}+X (X+2)-1\right)+X^4}}\times \nonumber \\
&&\left(\frac{\tilde{a}^2 \left(-\alpha ^2+(\alpha +1)^2 X^4+4 \alpha  (\alpha +1) X\right)+(\alpha +1) X^4 (\alpha +(\alpha +1) X(X-2))}{(\alpha +1) X^2 \left(\tilde{a}^2 ((\alpha +1) X(X+2)-\alpha )+(\alpha +1) X^4\right)}\right)^{-1/2},
\end{eqnarray}
and putting $\Omega_K=  \frac{U^{\phi}}{U^{t}}$ from (\ref{tdot1}), and (\ref{pdot1}), after some algebraic manipulation, one can obtain co--rotating Keplerian velocity in the ZAMO frame. Because the final expression is very long, we refrain from writing it here.
Henceforth, the dimensionless parameter $X=r/\mathcal{M}$, with a range $1<X<2$, indicates  X-point reconnection radial location. Note that the range $1<X<2$, in essence, covers the region inside of the ergosphere. It means that $X=1$, and $X=2$ denote the close of event horizon, and the outer boundary of the ergosphere, respectively. By taking the above equation into account in the expression (\ref{ex}), one can conclude that the energy-at-infinity per enthalpy of the plasma related to the accelerated/decelerated ($\pm$) plasma,  can be controlled via five parameters ($\tilde{a}$, $\alpha$, $X$, $\xi$, $\sigma_0$).
The origin of these parameters, in essence, comes from the nature of the black hole spacetime, and its environment. Concerning orientation angle $\xi$, a notable point is that there is no definitive agreement on its value, and it depends on various factors of the magnetic field configuration. Consistent with global numerical simulations of accretion flows onto a black hole, it is expected that the orientation angle is likely to vary in time, with intermittent fluctuations around some characteristic value. In light of some of the simulations such as Ref. \cite{Semenov:2004ib} (see also \cite{Bransgrove:2021heo}), the value of $\xi$ is expected to be in the range $[\pi/12,\pi/6]$. In this regard, one can also mention the simulations performed in \cite{McKinney:2012vh}, which indicate typical values of the orientation angle around $\pi/6$. As a result, throughout this manuscript, we consider the configuration of the magnetic field near a black hole in a manner where $\xi$ lies in the interval $[\pi/12,\pi/6]$.

First, let us start with showing the behavior of $\epsilon_{\pm}$ in terms of the orientation angle $\xi$, and the plasma magnetization $\sigma_0$ for an extremal rotating black hole, in Fig. (\ref{e1}).  The left panel displays the impacts of the orientation angle $\xi$ in interplay with $\alpha$ on $\epsilon_{\pm}^{\infty}$. Following the standard Kerr case, we also observe that increasing $\xi$ results in a decrease/increase in $\epsilon_{\pm}^{\infty}$, respectively.
As mentioned earlier, for energy extraction it is necessary that $\epsilon_{+}^{\infty}>0$, and $\epsilon_{-}^{\infty}<0$, meaning that the MOG parameter causes the upper limit expected from Kerr for the orientation angle $\xi$ to shift to larger values.
Therefore, if the Kerr--MOG black hole plays the role of the host for CAP, more values of $\xi$ can result in energy extraction i.e., beyond those allowed in the standard Kerr.
The right panel of Fig. (\ref{e1}) displays the impacts of the plasma magnetization $\sigma_0$ in interplay with $\alpha$ on $\epsilon_{\pm}^{\infty}$. First of all, in agreement with Kerr, we see that $\epsilon_{\pm}^{\infty}$ increases/decreases, respectively, as $\sigma_0$ increases.
Taking the MOG parameter in the spacetime of a rotating black hole shifts the lower bound of $\sigma_ 0$ required for energy extraction to less than expected from Kerr i.e., $\sigma_ 0>1/3$. In other words, in the framework of the Kerr--MOG black hole, extracting rotational energy via CAP is possible even for values of plasma magnetization $\sigma_0$ lower than $1/3$.

In this regard, by providing an analysis of some parameter spaces, we survey the viability of extracting energy from rotating black holes via CAP. In Figs. (\ref{e2}), and (\ref{e3}), we respectively display the regions of the parameter spaces $\tilde{a}-X$, and $\alpha-X$ which fulfill energy extraction conditions $\epsilon_{-}<0$, and $\epsilon_{+}>0$. Concerning Fig. (\ref{e2}), for both black holes (Kerr, and Kerr--MOG) we see that the orientation angle affects the parameter space $\tilde{a}-X$ so that it becomes more limited by increasing $\xi$.  By comparing panels in the top and bottom rows, one can deduce that the presence of the MOG parameter in the background of the Kerr black hole causes the parameter space to become wider. It is more evident in cases of near--extremal rotation and low plasma magnetization.  Based on Fig. (\ref{e3}) we first see that the increase of $\xi$ causes the parameter space  $\alpha-X$ to become compact.  Another important output from these panels in Fig. (\ref{e3}) is that increase in the plasma magnetization lets the MOG parameter for interplay with more values of $X$.

\section{Power and efficiency of the CAP from Kerr--MOG black hole}\label{EF}
We conclude our analysis by evaluating the power and efficiency of energy extraction through the CAP hosted by a rapidly spinning Kerr--MOG black hole. These two quantities depend on the amount of plasma with negative energy-at-infinity that is swallowed by the black hole in the unit time.  Expressed in mathematical terms,  the power $P_{\text{CAP}}$ per unit of enthalpy harvested from the rotating black hole via the escaping plasma can be written as \cite{Comisso:2020ykg}
\begin{eqnarray}
\mathcal{P}_{\text{CAP}}=\frac{P_{\text{CAP}}}{w_0}=-\epsilon_{-}^{\infty}A_{\text{in}}U_{\text{in}},~~~~A_{\text{in}}\sim(r_{\text{SLS}}^2-r_{\text{ps}}^2) \, ,
\end{eqnarray}
where $A_{\text{in}}$ is the cross sectional area of the inflowing plasma, which is $\sim3M$ for the Kerr black hole. $U_{in}$ is a constant parameter that depending on the type of plasma i.e., collisionless and collisional typically sets values $\mathcal{O}(10^{-1})$, and $\mathcal{O}(10^{-2})$, respectively \cite{Comisso:2016ima}.
By setting value  $U_{\text{in}}=\mathcal{O}(10^{-1})$, for the plasma near black hole i.e., collisionless, and employing $r_{\text{SLS}}$ from (\ref{sls1}), along with taking $r_{\text{ps}}\sim M$,  one can plot $\mathcal{P}_{\text{CAP}}$ for a rapidly spinning black hole in terms of the dominant X--point location, and the plasma magnetization $\sigma_0$, as shown in Fig. (\ref{e4}).  First, we observe that the presence of the MOG parameter causes grows of power. Second, by increasing $\alpha$ the peaks of power shift to smaller values of X--point location.  This implies that the Kerr--MOG black hole amplifies the power of energy extraction by shifting the location of the magnetic reconnection towards the event horizon. Third, the orientation angle $\xi$ affects the behavior of power in terms of $X$, and $\sigma_ 0$ so that by increasing it, the power drops.

Another important factor that can help assess the feasibility of energy extraction via the CAP from a Kerr--MOG black hole is the efficiency of the plasma energization process. It is defined as
\cite{Comisso:2020ykg}
\begin{eqnarray}\label{44}
\eta_{\text{CAP}}=\frac{\epsilon_{+}^{\infty}}{\epsilon_{+}^{\infty}+\epsilon_{-}^{\infty}}.
\end{eqnarray}
If the efficiency $\eta_{\text{CAP}}$ is greater than 1, it indicates that energy will be extracted from the rotating black hole. As is evident from  Fig. (\ref{e5}),  smaller values of X are more likely to result in $\eta_{\text{CAP}}>1$. Besides, in the presence of the MOG parameter, the dominant X--point locations that yield $\eta_{\text{CAP}}>1$ are larger than the one predicted for a Kerr black hole. Concerning the plasma magnetization this happens for smaller values of $\sigma_0$ expected from Kerr. In Fig. (\ref{e6}),  we show the plot of $\eta_{\text{CAP}}-\alpha$ for different values of $X$, $\sigma_0$. We see for a Kerr--MOG black hole surrounded by plasma with low magnetization as $\sigma_0=1$, the efficiency of the plasma energization process becomes $\eta_{\text{CAP}}>1$ provided that the dominant X--point location happens nearer to the event horizon i.e., $X=1.2$, $1.3$.  As the plasma magnetization increases, this requirement becomes less stringent.
As the final word here, in \cite{Comisso:2020ykg} shown that by adopting optimal conditions $X\rightarrow1$, $\xi\rightarrow0$, and $\sigma_0\gg1$ the maximum
efficiency for a near extremal rotating black hole is $\eta_{\text{CAP(max)}}\simeq3/2$. By repeating this condition for the host black hole at hand, Fig.  (\ref{ee6}) explicitly shows that in the presence of the MOG parameter we dealing with maximum efficiency bigger than that in the Kerr case. The root of the observed deviation in comparison to the standard Kerr, in essence, comes from the MOG parameter $\alpha$. To be precise, the effect of $\alpha$ on the behavior of $\epsilon_{\mp}^{\infty}$ is the reason the energy extraction efficiency of the Kerr-MOG black hole is higher than Kerr. Based on the definition presented in Eq. (\ref{44}) and the quantitative behavior of $\epsilon_{\mp}^{\infty}$ in Fig. (\ref{e1}) (right panel), one can realize that in the presence of the MOG parameter, the expression of $\epsilon_{+}^{\infty}+\epsilon_{-}^{\infty}$ in the denominator of Eq. (\ref{44}) becomes smaller than the case of Kerr. Namely, the efficiency of the Kerr-MOG black hole is higher than its standard counterpart.

\begin{figure}[ht!]
\begin{tabular}{c}
        \includegraphics[width=.48\linewidth]{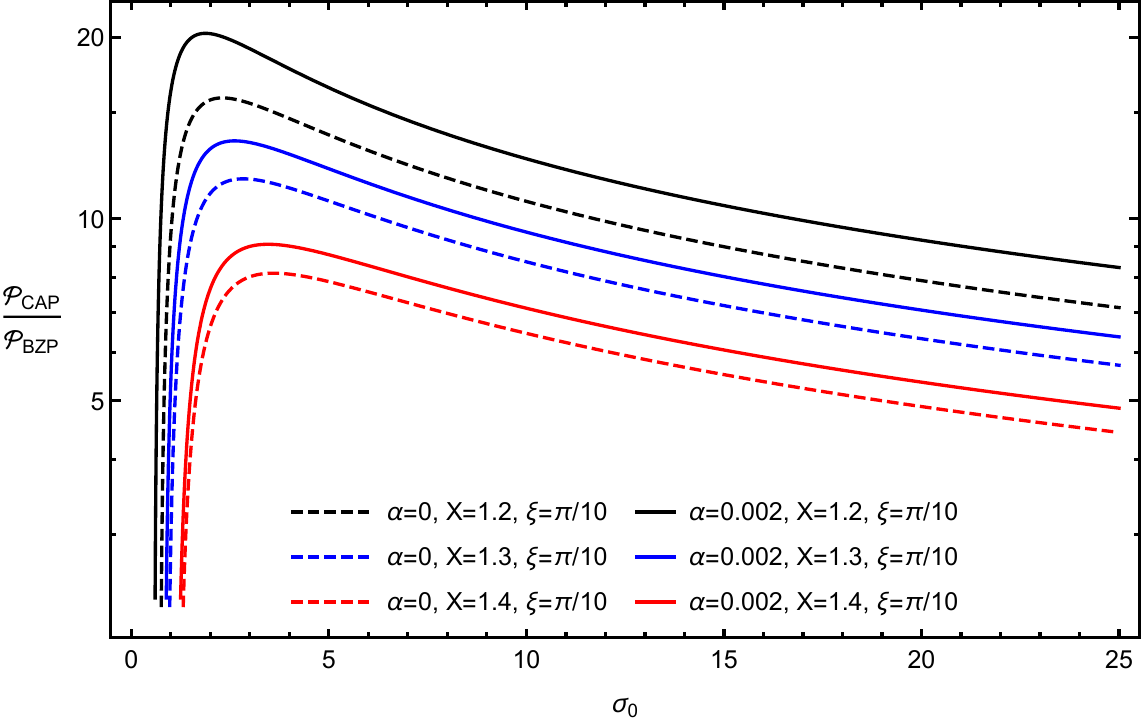}~~~
        \includegraphics[width=.48\linewidth]{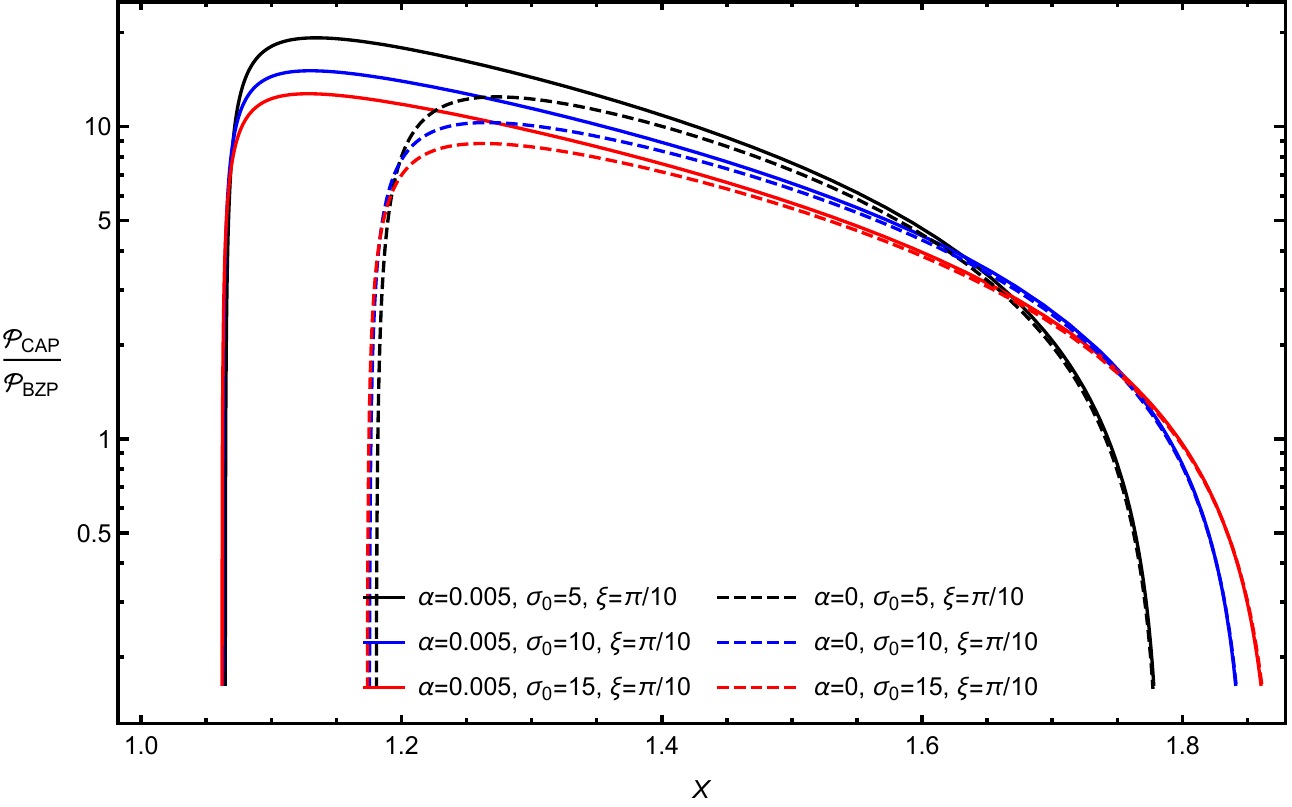}\\
    \end{tabular}
    \caption{Power ratio $\frac{\mathcal{P}_{\text{CAP}}}{\mathcal{P}_{\text{BZP}}}$ in terms of the plasma magnetization $\sigma_0$ (left panel), and the dominant magnetic reconnection X-point (right panel). Values $1.38,~-9.2$, and $0.044$ are used for the coefficients $c_{1,2}$, and $\kappa$, respectively. } \label{e7}
\end{figure}

\begin{figure}[ht!]
    \centering
    \begin{tabular}{c}
        \includegraphics[width=.6\linewidth]{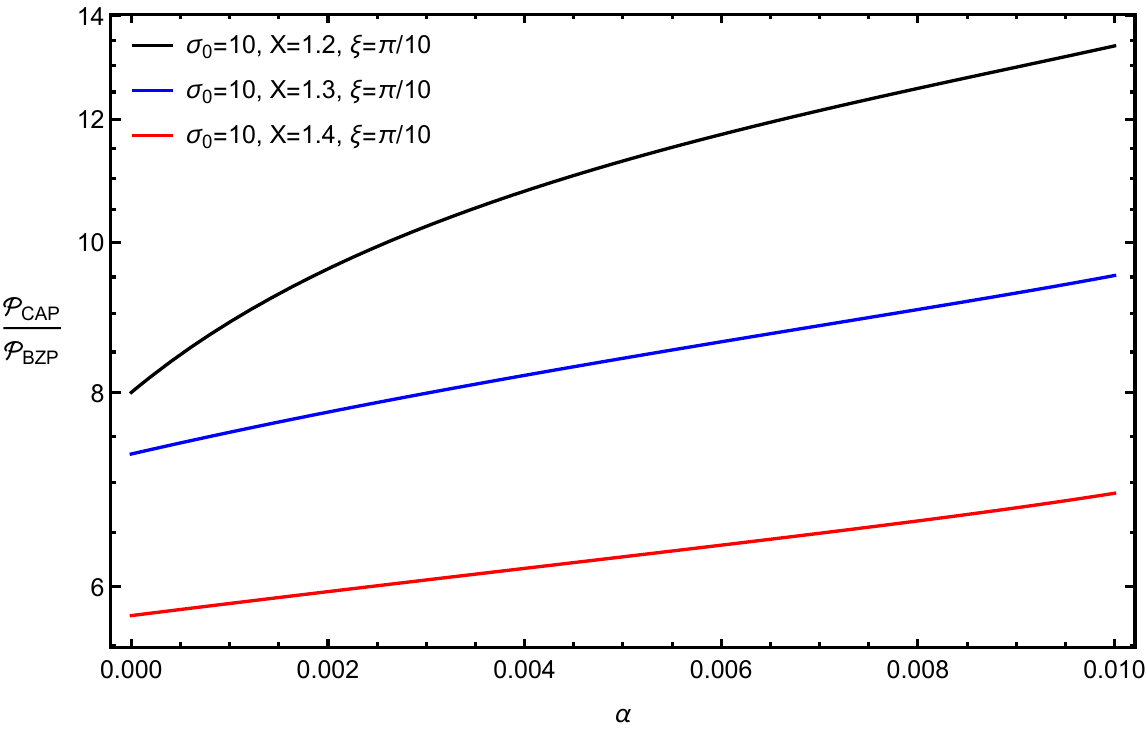}~~~
    \end{tabular}
    \caption{ Power ratio $\frac{\mathcal{P}_{\text{CAP}}}{\mathcal{P}_{\text{BZP}}}$  in terms of the MOG parameter $\alpha$.} \label{e8}
\end{figure}
\begin{figure}[ht!]
    \centering
    \begin{tabular}{c}
        \includegraphics[width=.47\linewidth]{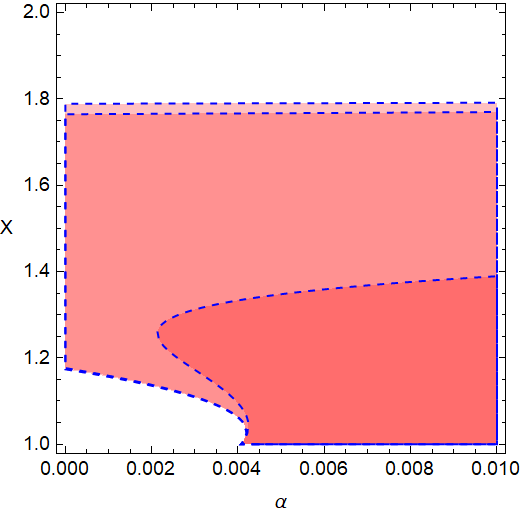}~~~
        \includegraphics[width=.47\linewidth]{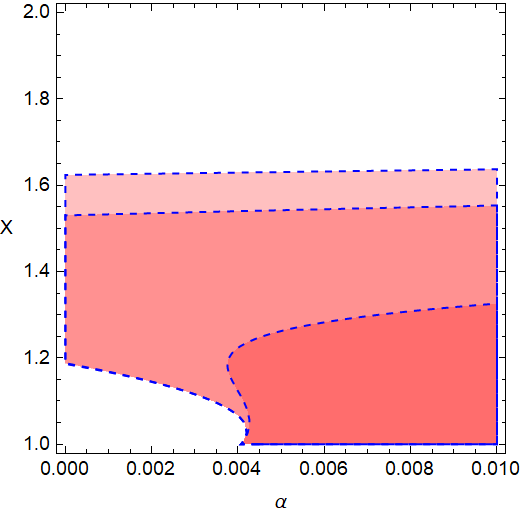}
    \end{tabular}
    \caption{Regions of the parameter space $\alpha-X$ in which $\frac{\mathcal{P}_{\text{CAP}}}{\mathcal{P}_{\text{BZP}}}>1$. The regions shaded in red (from dark to light) respectively belong to the plasma magnetization $\sigma_0\in \{1,10,50\}$ with orientation angles $\xi=\pi/10$ (left), and $\xi=\pi/8$ (right). Borderlines (blue-dashed) separate these areas.} \label{e9}
\end{figure}

\subsection{Comparison with the Blandford--Znajek process (BZP)}
We are now in a position to compare the power extracted from the black hole through the CAP with that extracted through BZP. The magnetic field, in essence, is the key component of both energy--harvesting processes.  However, in the BZP, the magnetic field plays a different role,  as it enables energy extraction by inducing an electric potential difference. This occurs through the externally generated magnetic field lines threading the black hole via the currents flowing in an equatorial disc  \cite{Blandford:1977ds}.
It has already been shown in \cite{Comisso:2020ykg} for a rapidly rotating host black hole that the CAP can be more efficient than the BZP.  In light of this comparison, we will enable to determine the impact of the MOG parameter on the previous result, specifically how its presence in the background of the Kerr black hole influences the power extraction of both processes.

Based on the BZP \cite{Blandford:1977ds,Tchekhovskoy:2009ba}, the rate of energy extraction can be expressed by taking the leading order terms of the ZAMO's angular velocity $\Omega_{\text{EH}}$ of the event horizon,  which is written as
\begin{eqnarray}\label{po}
\mathcal{P}_{\text{BZP}}=\frac{\kappa}{16\pi} \Phi^2_{\text{BH}} \Omega_{\text{EH}}^2
\left(1+c_1 \Omega_{\text{EH}}^2+c_2\Omega_{\text{EH}}^4+
\mathcal{O}(\Omega_{\text{EH}}^6)\right),
\end{eqnarray}
where $c_{1,2}$ are numerical coefficients with approximate values $1.38$ and $-9.2$,  respectively,  and a numerical constant $\kappa\approx0.044$ that comes from the magnetic field configuration \cite{Pei:2016kka}.
In the above expression, $\Phi_{\text{BH}}$ is the magnetic flux threading one hemisphere of the black hole horizon and is defined as $\Phi_{\text{BH}}=\frac{1}{2}\int_{\theta}\int_{\phi}|B^r|dA_{\theta\phi}\sim B^{r}A_{EH}/2\sim 2\pi(r_{\text{EH}}^{2}+\tilde{a}^2)B_0\sin\xi$.
As a result,  the power ratio between the CAP and the BZP can be expressed as
\begin{eqnarray}\label{46}
    \frac{\mathcal{P}_{\text{CAP}}}{\mathcal{P}_{\text{BZP}}}=
    \frac{-4\epsilon_{-}^{\infty}A_{\text{in}}U_{\text{in}}}
    {\pi\kappa\sigma_0(\tilde{a}^2+r_{EH}^2)^2 \Omega_{\text{EH}}^2 \sin^2\xi(1+c_1 \Omega_{\text{EH}}^2+c_2\Omega_{\text{EH}}^4)}.
\end{eqnarray}
By substituting the angular velocity of the Kerr--MOG black hole, which is given by
\begin{eqnarray}
\Omega_{\text{EH}}=-\frac{g_{t\phi}}{g_{\phi\phi}}=\frac{2\tilde{a}r_{\text{EH}}}{\left(2r_{\text{EH}}+\frac{\alpha}{\alpha+1}\mathcal{M}\right)^2} ,
\end{eqnarray}
we can calculate the ratio of power between the CAP and the BZP. It would be helpful to comment here on the comparison of the power extracted from the black hole via CAP and BZP. The ratio appeared in Eq. (\ref{46}) is merely a rough estimate of the power extracted during fast magnetic reconnection relative to the approximately steady-state BZP. While the CAP operates efficiently through the azimuthal component of the magnetic field, the BZP operates efficiently through the radial component. More precisely, the magnetic configuration around the black hole is highly dynamic, and one can expect the equatorial plane to be in favor of the CAP, while the BZP should operate more efficiently at higher latitudes close to the poles. Due to the uncertainties and the time dependence of the magnetic configuration around the black hole, there are significant uncertainties so that global numerical simulations are needed to fully evaluate this issue. In this regard, the principle assumption utilized in this comparison is using an approximation in the form of $\Phi_{\text{BH}}\sim B^{r}A_{EH}/2\sim 2\pi(r_{\text{EH}}^{2}+\tilde{a}^2)B_0\sin\xi$. However, an accurate evaluation of $\Phi_{\text{BH}}$ requires direct numerical simulations that exactly reproduce the magnetic field configuration at all latitudes. While the orientation angle $\xi$ is merely a good estimate for the magnetic field configuration at low latitudes. It is emphasized that this merely works from the point of view of a rough estimate, and to accurately calculate the power of BZP, the use of numerical simulations is necessary since the field lines that thread the event horizon around the poles might have a quite different orientation than what is happening in the equatorial plane. For this reason, to accurately calculate the power of BZP, it is essential to know the direction of the magnetic field at all latitudes and not only near the equatorial plane.
Another important point that Eq. (\ref{46}) reveals to us is that in limit $\sigma_0\rightarrow \infty$ (or $B_0\rightarrow \infty$) then energy extraction via CAP is always subdominant to the BZP since $\frac{\mathcal{P}_{\text{CAP}}}{\mathcal{P}_{\text{BZP}}}\rightarrow0$. All in all, these indicate that the result of this comparison depends on magnetic field configuration as well as the strength of the magnetic field.

The results are shown in Figs. (\ref{e7}) and (\ref{e8}) for different conditions.
Both panels in Fig. (\ref{e7}) reveal that in the presence of the MOG parameter $\alpha$, the ratio between the powers of the CAP  and BZP is greater than in the case of $\alpha=0$ (Kerr). Moreover, in the Kerr--MOG black hole, the peak of this ratio occurs at the magnetic reconnection X--points closer to the event horizon, as shown in the right panel.  For a better understanding, one can see Fig. (\ref{e8}) which shows the growth of this ratio as $\alpha$ increases. Consistent with previous results, we observe that the closer the magnetic reconnection X--point occurs near the event horizon, the greater the growth of this power ratio.  In order to evaluate the parameters for which the CAP is more efficient than the BZP, we require the condition $\frac{\mathcal{P}_{\text{CAP}}}{\mathcal{P}_{\text{BZP}}}>1$ to hold within the parameter space $\tilde{a}-X$ for different values of the plasma magnetization $\sigma_0$ and orientation angle $\xi$.  As can be seen in Fig. (\ref{e9}), increasing $\sigma_0$ leads to an enlargement of the parameter space $\tilde{a}-X$ satisfying the condition $\frac{\mathcal{P}{\text{CAP}}}{\mathcal{P}{\text{BZP}}}>1$.  Furthermore, a comparison between the left and right panels reveals that the orientation angle has an effect on the allowed range of the parameter space $\tilde{a}-X$, such that decreasing the orientation angle widens the parameter space toward a higher X--point magnetic reconnection location.  In summary, these results suggest that the energy extraction via the CAP can exceed that of the BZP, making it a more efficient mechanism for extracting rotational energy from black holes.

\section{Summary and conclusions}\label{Co}
It is now well established that magnetic reconnection can occur in the equatorial plane of a rotating black hole provide that the plasmoid instability triggers rapid changes in magnetic field connectivity.
This phenomenon is highly relevant for modeling the extraction of
rotational energy from black holes. In this paper, we have considered the equatorial plane of a Kerr--MOG black hole as a host for energy extraction through the magnetic reconnection recently proposed by Comisso and Asenjo \cite{Comisso:2020ykg}.
The underlying energy extraction model that we have labeled
Comisso--Asenjo process (CAP) is based on the occurrence of fast
magnetic reconnection in the ergosphere of an extremal rotating
black hole. Unlike the Penrose process (PP), which extracts energy using mechanical particle fission i.e., producing two particles with negative and positive energy-at-infinity, the CAP involves the reconnection of magnetic field lines, which generates plasma flows with both negative and positive energy-at-infinity.
This plasma flows extract energy from the black hole, as the
escaping plasma particles carry extra energy, as the conservation
of energy dictates.

Owing to the importance of the ergosphere in this model as the
location for occurring magnetic reconnection and subsequently plasma acceleration, we have first discussed the radii of
the event horizon, stationary limit surface (SLS), and circular
photon orbit for a Kerr--MOG black hole. Concerning the last one,
the co--rotating orbit drew our attention since just in this case
the occurrence of magnetic reconnection inside the ergosphere is
guaranteed.  The remarkable thing about the MOG parameter $\alpha$
is that it obeys different upper bounds due to the dependence on
the mass of the gravitational central sources.  Given the aim of
this paper, namely, modeling energy extraction from supermassive
black holes via the CAP,  throughout our analysis we have set
values $\alpha \lesssim 0.01$. This constraint, in essence, is the
most stringent bound on the MOG parameter derived recently from
the observation of a supermassive black hole's shadow Sgr A* in the
center of Milk Way galaxy \cite{Vagnozzi:2022moj}. 
In light of this constraint, the existing Kerr--MOG black hole is near extremal with dimensionless rotation parameter
$\tilde{a}\sim0.99$.

Starting point of our analysis, in essence, is reproduce of the expression of energy-at-infinity ($\epsilon_{\pm}^{\infty}$) for the accelerated/decelerated plasma (positive/negative sign) \cite{Comisso:2020ykg}, in the framework of Kerr--MOG spacetime. The presence of the parameter $\alpha$ in addition to the standard parameters ($\tilde{a}$, $X$, $\sigma_0$, $\xi$) makes the new framework richer. 
To satisfy the energy extraction conditions ($\epsilon_{+}^{\infty}>0$, and $\epsilon_{-}^{\infty}<0$), we found that in the presence of the MOG parameter, the upper limit of the orientation angle $\xi$ moves to larger values (compared to Kerr black hole). 
Namely, if a Kerr--MOG black hole hosts the CAP, there are more values of the orientation angle $\xi$ that result in energy extraction compared to a standard Kerr black hole.
Apart from
increasing/decreasing $\epsilon_{\pm}^{\infty}$ by increasing
$\sigma_0$,  the presence of the MOG parameter in the spacetime of
a rotating black hole shifts the lower bound of $\sigma_ 0$
required for energy extraction to lower values than what is
expected from Kerr black holes,  i.e., $\sigma_ 0>1/3$.
Specifically, in the MOG black holes, energy can be extracted for
plasma magnetization values below $\sigma_0 < 1/3$, too. These
explanations can be traced in Fig. (\ref{e1}). By applying the
energy extraction conditions $\epsilon_{+}^{\infty}>0$, and
$\epsilon_{-}^{\infty}<0$, we have depicted some allowed regions
of parameter spaces $\tilde{a}-X$, and $\alpha-X$ in Figs.
(\ref{e2}), and (\ref{e3}), respectively. In this way, one can
find the allowed range resulting interplay of the involved
parameters via certain parameter regions.

By considering a rapidly spinning MOG--black hole as a host
spacetime for the CAP,  we have computed the power and efficiency
extracted from the black hole.  For the dominant magnetic
reconnection X--points close to the event horizon, the power
extracted from the black hole is larger than that in a Kerr black
hole.  As smaller and larger values are set for the orientation angle $\xi$ and the MOG parameter $\alpha$, the power becomes larger.
Concerning the efficiency of the plasma energization process, it is found that it increases with $\alpha$, and $\sigma_0$. A key factor in energy extraction is achieving an
efficiency greater than 1, which is denoted by
$\eta_{\text{CAP}}>1$. As $\alpha$ increases, the entry point into
the region $\eta_{\text{CAP}}>1$ occurs for smaller and larger
values of $\sigma_0$ and X, respectively. More details can be
found in Figs. (\ref{e4})--(\ref{e6}). 
By taking the optimal
conditions $X\rightarrow1$, $\xi\rightarrow0$, and $\sigma_0\gg1$ into account, for a near extremal rotating black hole we in the Fig. (\ref{ee6}) have shown that in the presence of the MOG parameter, the maximum efficiency ($\eta_{\text{CAP(max)}}$) is
bigger than that in Kerr i.e., $\eta_{\text{CAP(max)}}>3/2$.

We concluded our study by comparing the power of the CAP, and the
BZP via numerical analysis of this ratio in Figs.
(\ref{e7})--(\ref{e9}).  An important observation is that the peak of the mentioned power ratio appears at relatively low values of $\sigma_0$ when the dominant X-point location approaches the event horizon. 
This ratio increases with the MOG
parameter so that for cases with X near the event horizon it is
bigger than cases with X away from the event horizon.  In general,
the existence of $\alpha$ directs this ratio to larger values than
the case of $\alpha=0$. In particular, in certain regions of
parameter space $\alpha-X$, this ratio is higher than one indicating that
the CAP is more efficient than the BZP.  This implies that the CAP
hosted by Kerr--MOG black hole can serve as a more practical energy
extraction model from the Kerr black hole.

In summary, the MOG parameter in the rotating black hole enhances
the power and efficiency of energy extraction via the CAP, as
compared to the standard Kerr black hole.  
The presence of the MOG
parameter in the spacetime background can make the CAP more
efficient than the BZP. As a final word, this paper presents a promising energy extraction via CAP hosted by a rapidly rotating MOG black hole, leading to more energy extraction from the Kerr black hole.
\\

\vspace{1cm} {\bf Acknowledgments:} We thank Luca Comisso for
carefully reading the manuscript and for valuable discussions and
comments. DFM thanks the Research Council of Norway for their
support and the UNINETT Sigma2 -- the National Infrastructure for
High Performance Computing and Data Storage in Norway.


\end{document}